\documentclass[breakmath]{jsedi_pub}
\usepackage[dvipsnames]{xcolor} %added package 
\usepackage{dynamo_symbols}
\usepackage{overpic}
\usepackage{hyperref}
\usepackage{xr-hyper}   
\usepackage{comment}
\usepackage{overpic}
\usepackage{lineno}
\usepackage{csvsimple}
\usepackage{longtable} % Make sure this is in your preamble
\usepackage{xr}
\newif\ifarxiv
 \arxivtrue  % enable before arXiv upload
\newcommand{\figfile}[1]{%
\ifarxiv
  #1-eps-converted-to.pdf%
\else
  figures/#1.eps%
\fi
}

\title{Thermochemical models of outer core convection with heterogeneous core-mantle boundary heat flux}
\shorttitle{Thermochemical convection with core-mantle boundary heterogeneity} % used for header, not mandatory but recommended

\author[1]{Souvik Naskar
	\orcid{0000-0003-0445-8417}
	\thanks{Corresponding author: 
\href{mailto:s.naskar@leeds.ac.uk}{\texttt{s.naskar@leeds.ac.uk}}}
}
\author[1]{Jonathan E. Mound
	\orcid{0000-0002-1243-6915}
}
\author[1]{Christopher J. Davies
	\orcid{0000-0002-1074-3815}
}
\author[1]{Andrew T. Clarke
	\orcid{0000-0003-2128-0016}
}
\affil[1]{School of Earth and Environment, University of Leeds, Woodhouse Lane, Leeds, LS2 9JT, United Kingdom}
\credit{Conceptualisation}{Souvik Naskar, Jonathan E. Mound, Christopher J. Davies}
\credit{Software}{Andrew T. Clarke}
\credit{Formal Analysis}{Souvik Naskar, Jonathan E. Mound, Christopher J. Davies}
\credit{Resources}{Souvik Naskar, Andrew T. Clarke}
\credit{Writing - Original draft}{Souvik Naskar}
\credit{Writing - Review \& Editing}{Souvik Naskar, Jonathan E. Mound, Christopher J. Davies}
\begin{document}

%%%-- for publication only, using data (doi, etc.) provided by the editorial board upon acceptance of paper 
\publicationonly{
\dois{10.46298/jsedi.17084}
\handedname{Mathieu Dumberry}
\receiveddate{December 11, 2025}
\reviseddate{ April 18, 2026; May 8, 2026}
\accepteddate{May 9, 2026}
\publisheddate{May 15, 2026}
\theyear{2026}
\thevolume{2}
\thepaper{4}  %% article number
}

%% Your article can include up to 3 abstracts. The first is the English language abstract.
%% For other languages in the second and third optional abstracts, you might have to define 
%% additional font(s) in preamble above
%% You can also include a non-technical summary in addition to the abstract(s)
\makesedititle{
\begin{summary}{Abstract}
Convection in Earth's outer core is driven by the release of heat and light elements at the inner core boundary. A key question is whether these buoyancy sources drive convection throughout the core, or whether a stable layer exists just below the core-mantle boundary (CMB). Recent simulations incorporating CMB heat flux heterogeneities propose locally stable ``regional inversion lenses'' (RILs) rather than a global layer, allowing stable and unstable regions to coexist. However, these simulations combine thermal and compositional anomalies, ignoring differences in diffusivities and boundary conditions. Here we simulate thermal, chemical, and thermochemical convection at Ekman number $E=10^{-5}$, with thermal and chemical flux Rayleigh numbers $\tRaT=30-4000$ and $\tRaC=30-100000$, and Prandtl numbers $Pr_T=1$ and $Pr_\xi=10$. Purely chemical simulations accumulate light elements below the CMB, forming locally stable regions near the poles or global layers, depending on $\tRaC$. These chemically stratified regions persist in thermochemical simulations even when thermal forcing is destabilising. Introducing heterogeneous CMB heat flux produces thermally stratified RILs even with strongly destabilising compositional buoyancy. Our simulations reveal a diverse range of locations, properties, and morphologies of stable regions depending on $\tRaT$ and $\tRaC$, they can have seismically detectable thickness and strength, and might also have a signature in geomagnetic observations.
%\clearpage
\end{summary}
\begin{summary}{Non-technical summary}
%\begin{comment}
Earth’s magnetic field is generated by fluid motion in the liquid iron outer core, which arises from the release of heat and light elements at the inner core boundary. A long‑standing question is whether fluid motion arises throughout the core, or whether parts of the uppermost core, just beneath the core-mantle boundary (CMB), resist fluid motion and are therefore stably stratified. The existence of stable regions at the top of Earth's core has the potential to strongly affect how we interpret seismic and geomagnetic observations, but their existence, properties, and origin remain uncertain. Here, we use numerical simulations of core dynamics to understand the formation and properties of stable regions below Earth's CMB. Unlike previous studies, we consider core fluid flow driven both by the release of heat and light elements at the inner core boundary and by the presence of thermal anomalies in the overlying mantle. We show that light elements can naturally accumulate beneath the CMB, particularly in polar regions, forming chemically stable regions even when thermal buoyancy is strong. Thermally stable regions form in equatorial regions below Africa and the Pacific, where the lower mantle is anomalously hot, and persist despite strong chemical buoyancy. These stable regions vary widely in size, strength, and location, can be thick enough to be detected seismically, and may influence observable features of Earth's magnetic field.
%\vspace{1.3cm} % because abstract is too short
%\end{comment}
\end{summary}
}

\section{Introduction}\label{sec:intro}
%% Labels are used to cross-reference an item using \ref command.
%Context/background/Problem
The core-mantle boundary (CMB) remains one of the most enigmatic regions of the deep interior of our planet, and a decades-old debate continues over whether the uppermost core is convecting or stably stratified. The presence of local regions or a global layer of stability would fundamentally change interpretations of geomagnetic observations, which only probe the top of the core, by obscuring the underlying convection and dynamo action, and may have important geophysical implications for the long-term thermochemical evolution of the core \citep[see e.g.,][]{buffett2010stratification,brodholt_2017,greenwood2021evolution, davies_2023}. Regions that depart from a well-mixed isentropic profile could be imaged seismically if there is an accompanying departure of the compressional wave velocity; however, previous studies (which generally interpret velocity anomalies as a globally stable layer) have advocated for a range of scenarios, from no seismically visible layer \citep{alexandrakis_2010,irving2018seismically}, to layers several hundreds of kilometres thick \citep{kaneshima_2018,helffrich_2010}. A layer that is stable to convection could also affect the geomagnetic secular variation by suppressing radial motion, and the presence or absence of radial flow near the top of the core is also a long-standing topic of debate \citep{whaler_1980, whaler_1986, Amit2014CanVariation, lesur_2015,huguet_2018}. Other indirect observational constraints on core stratification rely on specific mechanisms: \citet{gubbins_2007} argued that a stable layer could be at most 100~km-thick if flux expulsion is the cause of the current reversed flux enhancement in the southern hemisphere, while \citet{buffett_2014} argued for a $\orderof{100}$~km-thick layer in order to explain a $60$-yr dipole fluctuation by a certain class of magnetohydrodynamic waves. The present uncertainty regarding the stability of the uppermost core restricts our ability to predict the heat and mass flux through the CMB, which are vital for elucidating the structure, dynamics, and evolution of the CMB region. 

%Previous Studies on stratification
The origin of a global stable layer below the CMB is generally attributed to mechanisms external to the core, such as incomplete mixing during core formation \citep{landeau_2016}, subadiabatic heat flow at the CMB \citep{gubbins1982stable, labrosse1997cooling, Pozzo2012ThermalConditions, greenwood2021evolution,labrosse_2015} or chemical interactions with the mantle \citep[e.g.,][]{buffett2010stratification, brodholt_2017, davies_2023}. Numerical studies generally assume a global layer \citep{christensen2006deep,nakagawa2011effect,christensen_2018} and find strongly suppressed radial motion below the CMB and surface magnetic fields that become morphologically dissimilar to the observed geomagnetic field as the strength and thickness of the layer are increased \citep{gastine_2020,aubert_2025}. However, flow and field behaviour can change significantly when simulations incorporate lateral variations in CMB heat flow. \citet{olson_2017} found that strong lateral heat flow variations can partially destabilise a pre-existing stable layer, producing localised regions of radial flow. \citet{mound_2019} studied non-magnetic rotating convection with no pre-existing layer and found that stable regions at the top of the core can arise by imposing a laterally varying CMB heat flux pattern derived from seismic tomography. These $\orderof{100}$~km-thick stable regions, called Regional Inversion Lenses (RILs), arise under the Pacific and African Large Low Velocity Provinces (LLVPs) where heat flow is suppressed and are characterised by a stabilising temperature gradient and weak upwelling flow. These stable regions offer the possibility to reconcile the conflicting interpretations of seismic and geomagnetic observations by allowing the coexistence of both stable regions with suppressed radial flow and unstable regions without any such flow suppression. \citet{mound_2023} found that dynamo simulations with dynamically generated RILs match the modern-day morphology and secular variation of the geomagnetic field. 

%Gap in literature
Previous studies of stable regions have considered convection driven by thermal buoyancy \citep{mound_2019} or a generalised ``codensity'' that combines the effects of latent heat and light element release due to solidification of the inner core \citep{olson_2017}. While these approaches are theoretically and computationally convenient, they ignore double-diffusive effects arising from the vastly different diffusivities and boundary conditions of the thermal and chemical fields. The purpose of this study is to understand how these effects influence the structure and stability of RILs.  

%Details of our model
Various regimes of double diffusive convection can arise depending on the relative size of the diffusion coefficients and the component (temperature or composition) that is stabilising \citep[e.g.,][]{turner1973buoyancy}. In this study, we focus on the ``top-heavy'' regime where both temperature and composition are destabilising, but have different diffusivities and boundary conditions. Few studies have considered this configuration in rotating spherical shell geometry \citep{breuer_2010,trumper_2012, tassin_2021} and none have imposed laterally varying outer boundary conditions on the thermal field. Here we use the ``tomographic'' heat flow pattern and thermal buoyancy profile employed in our previous studies \citep{mound_2017, mound_2019, mound_2023} in a suite of top-heavy double-diffusive rotating spherical shell convection simulations. The compositional field is destabilising throughout the core and can therefore potentially disrupt the RILs. On the other hand, previous studies have found that compositional stratification due to light-element accumulation (LEA) can arise naturally from the internal dynamics of chemical convection \citep{bouffard_2019}.

%Present study: Imporant results
In this study, we present $92$ new numerical models, with $7$ purely chemical and $85$ thermochemical top-heavy double-diffusive simulations of non-magnetic rotating spherical shell convection, combined with previous purely thermal models \citep{mound_2017}. In the thermochemical models, we consider the CMB heat flux to be either spatially homogeneous ($49$ models) or heterogeneous ($36$ models) with a pattern derived from seismic tomography (henceforth designated as homogeneous and heterogeneous simulations, respectively). Our objective is to explore the double-diffusive fluid dynamics in order to investigate the existence, morphology, strength, and thickness of the RILs. We find LEA-driven chemical stratification in both purely chemical and thermochemical homogeneous models. This reveals the existence of quasi-stationary stable regions/layers driven by LEA, for the first time, complementing previous work that identified non-stationary chemically stable layers in purely chemical rotating convection \citep{bouffard_2019}. In the heterogeneous thermochemical models, we find that LEA below CMB may lead to both local and global stable regions with similar strength and thickness as the RILs. However, the RILs in our thermochemical models remain the dominant stable regions near the topmost core in most heterogeneous models for the region of parameter space explored. The RILs in our new thermochemical simulations have similar strength and thickness to the previous pure thermal models \citep{mound_2019}. 

%Paper Organization
This paper is organised as follows. The numerical model is briefly described in section \ref{sec:methods}, with the full model equations detailed in Appendix  \ref{app:SI}. The results are presented in section \ref{sec:results}; we assess the global force balance in our models in section \ref{sec:force_regime}, which relates to the behaviour of stable regions that we analyse in section \ref{sec:stable_region}. The results are summarised, and their implications for Earth's core are discussed in section \ref{sec:conclusions}.  All simulation diagnostics can be found in the supplementary table (see Appendix \ref{app:diagnostics}).

\section{Methods}\label{sec:methods}
%Adapted from Mound and Davies 2017, Monville et al. 2019, Trumper et al. 2012 etc.
We employ a numerical model of rotating convection of a Boussinesq fluid. A spherical coordinate system $(r,\theta,\phi)$ is used to represent the domain bounded by the inner and outer boundaries, $r_i$ and $r_o$, respectively. The whole system rotates with a constant angular velocity $\boldsymbol{\Omega}=\Omega\boldsymbol{\hat{z}}$ about the vertical axis, and gravity varies linearly with radius. Velocity boundary conditions are no-slip on $r_i$ and $r_o$. 

The fluid is a mixture of light components dissolved in a comparatively heavy liquid (e.g., oxygen mixed in liquid iron in Earth's outer core). The relevant physical properties of the mixture are the kinematic viscosity, $\nu$, the thermal and chemical diffusivities, $\kappa_T$ and $\kappa_\xi$, and the coefficients of thermal and chemical expansion, $\alpha_T$ and $\alpha_\xi$. The thermal diffusivity is defined as $\kappa_T=k_T/\rho_o c_p$, where $k_T$ is the thermal conductivity, $\rho_o$ is the reference density, and $c_p$ is the specific heat capacity of the mixture. 

In our simulations, thermal convection is driven by basal heating, which represents the release of latent heat at the inner core boundary (ICB) of Earth's core, subject to a fixed flux on both boundaries. The conductive temperature gradient at the boundaries is expressed as $\boldsymbol{\nabla}T_{c}=-(\beta_{T}/r^{2})\boldsymbol{\hat{r}}$, where $\hat{r}$ is the unit vector in the radial direction, and $\beta_{T}$ is related to the ICB heat flow through $\boldsymbol{Q}_{T,i}=4\pi r_{i}^{2}(-k_{T}\boldsymbol{\nabla}T_{c})=4\pi k_{T} \beta_T \boldsymbol{\hat{r}}$. To model the release of light elements from the inner core boundary, we follow a standard setup \citep[e.g.,][]{kutzner2002stable} and impose fixed flux conditions at $r_i$ such that $\boldsymbol{\nabla}\xi_{c}=-(\beta_{\xi}/r_i^{2})\boldsymbol{\hat{r}}$. 
We assume zero compositional flux at $r_o$ (i.e., $\boldsymbol{{Q}}_{\xi,o}=\boldsymbol{0}$). To ensure stationary solutions, we assume that the flux from the inner core is balanced by a spatially homogeneous sink ($S_\xi$) that maintains the global balance of lighter elements \citep[see e.g.,][]{Kono2001DefinitionSimulation}.

The governing equations for the conservation of mass, momentum, energy, and chemical composition are standard and can be found in equations \ref{eqn:continuity_d}-\ref{eqn:composition_d} in Appendix~\ref{app:SI}. They are nondimensionalised using the shell gap ($h = r_o - r_i$) as the length scale, the viscous diffusion time ($h^2/\nu$) as the time scale, $\beta_T/h$ as the temperature scale, and $\beta_\xi/h$ as the scale of the chemical field (equations \ref{eqn:continuity_nd}-\ref{eqn:composition_nd} in Appendix~\ref{app:SI}). The non-dimensional numbers appearing in these equations are the Ekman number ($E$), thermal and chemical flux Rayleigh numbers ($Ra_T$ and $Ra_\xi$), and thermal and chemical Prandtl numbers ($Pr_T$ and $Pr_\xi$) defined as
%\begin{equation}
%\begin{split}
\begin{align}
E = \frac{\nu}{2\Omega h^2},\qquad 
Ra_T=\frac{g_o\alpha_T\beta_T h^2}{\nu \kappa_T r^{*}_{o}}, \qquad 
Ra_\xi=\frac{g_o\alpha_\xi\beta_\xi h^2}{\nu \kappa_{\xi} r^{*}_{o}}, &\qquad \notag\\
Pr_T=\frac{\nu}{\kappa_T}, \qquad
Pr_\xi=\frac{\nu}{\kappa_\xi}.
\end{align}
%\end{split}
\label{eqn:nd_parameters}
%\end{equation}
Additionally, modified flux Rayleigh numbers are used in this study; they relate to the flux Rayleigh numbers as $\tRaT=Ra_TE$ and $\tRaC=Ra_\xi E$.

In addition to homogeneous thermal boundary conditions, in some simulations we impose a laterally heterogeneous thermal flux at the CMB, following \citet{mound_2017}. The parameter space now also includes the pattern and amplitude of lateral variation in the CMB heat flux.  In this study, the pattern is fixed using an inferred heat flux pattern from seismic tomography following \citet{masters_1996}. The amplitude of heterogeneity is characterised as $\qstar=(q^{T}_{max}-q^{T}_{min})/q^{T}_{avg}$   where $q^{T}_{max}$,  $q^{T}_{min}$, and $q^{T}_{avg}$ are the maximum, minimum, and horizontally averaged heat flux through the CMB, respectively. We consider values of $\qstar\in\{2.3,\ 5,\ 10\}$, where $\qstar>2$ has been reported to produce RILs for pure thermal convection \citep{mound_2019}. The homogeneous models correspond to $\qstar=0$. Notably, the average heat flux $q^{T}_{avg}$ is the same for homogeneous and heterogeneous models, independent of $\qstar$.

%We represent the radial gradient of density in terms of the Brunt-V\"{a}is\"{a}l\"{a} (BV) frequency defined as,
The stable regions that emerge in our simulations are characterised by their thickness ($\delta$) and strength ($\N$) as estimated from the locally averaged radial profiles of the squared Brunt-V\"{a}is\"{a}l\"{a} (BV) frequency ($\Nsq$), defined as

\begin{equation}\label{eqn:Nsq_nd}  
    \Nsq = r^{*}E^2\left(\GrT\Trnd+\GrC\Crnd\right)\equiv \NTsq+\NCsq,
\end{equation}
where $\NTsq$ and $\NCsq$ are the BV frequencies estimated from the individual radial gradients of thermal and chemical fields. The derivation of equation \ref{eqn:Nsq_nd} and the local averaging procedure is outlined in section \ref{sec:diagnostics} in Appendix~\ref{app:SI}. We define the thickness of a stable region as the depth beneath the CMB of the zero crossing of the $\Nsq$ profile (i.e., radial location of neutral stability, $r_s$) and its strength as the square root of the maximum positive value of $\Nsq$, that is

\begin{align}\label{eqn:delta}      
    \delta=r_o-r_s\ \text{where}\ \left(\Nsqeq\right)_{r=r_s}=0,
\end{align}

\begin{equation}\label{eqn:N}      
    \N_{\rm max} = \sqrt{max\left\{\Nsqeq(r)\right\}}.
\end{equation}

\subsection{Force balance}\label{sec:force}
The global force balance in our simulations can be assessed from the ratio of various terms in the momentum equation (Appendix~\ref{app:SI}, equation~\ref{eqn:momentum_nd}). In this equation, the second and third terms on the left-hand side describe the Inertia ($I$) and Coriolis ($C$) forces, whereas the second and third terms on the right-hand side represent the thermal and chemical buoyancies (i.e., Archimedean forces, $A_{T}$ and $A_{\xi}$). In the study by \citet{naskar_2025}, the forces and curled forces are partitioned into mean (i.e., azimuthal average) and corresponding fluctuating parts (see also \citet[e.g.,][]{calkins_2021,nicoski_2024}), and we only consider the fluctuating part of the forces (indicated by a prime), giving
%\begin{equation}
%\begin{split}
\begin{align}
     \boldsymbol{I}' =(\boldsymbol{u^{'}}\cdot\boldsymbol{\nabla})\boldsymbol{u^{'}} , ~~~
     \boldsymbol{C}' =   \frac{1}{E}\left(\boldsymbol{\hat{z}}\times\boldsymbol{u^{'}}\right), & \notag \\
     \boldsymbol{A}'_{T} = \left(\frac{Ra_{T}}{Pr_T}\tT'\right)\boldsymbol{r} , ~~~
     \boldsymbol{A}'_{\xi} = \left(\frac{Ra_{\xi}}{Pr_\xi}\tC'\right)\boldsymbol{r}.  ~~~     
\end{align}
%\end{split}
%\end{equation}
All forces are integrated over the bulk fluid, which excludes regions of radial thickness $10$ times the Ekman layer depth adjacent to the upper and lower boundaries (defined based on the linear intersection method). The results are time-averaged, and subsequent notation implicitly includes both spatial and temporal averaging.

The metrics we use to classify the dynamical balances are based on the ratio of the magnitudes of the inertia and Coriolis forces, and their curls \citep{naskar_2025} that are, respectively,

\begin{equation}\label{eqn:FIC}
\begin{split}
\mathcal{F}_{I/C} = \frac{|\boldsymbol{I'}|}{|\boldsymbol{C'}|}, ~~~
\mathcal{CF}_{I/C} = \frac{|\boldsymbol{\nabla\times I'}|}{|\boldsymbol{\nabla\times C'}|}.      
\end{split}
\end{equation}
Additionally, we introduce a metric to estimate the relative importance of the two buoyancy forces,

\begin{equation}\label{eqn:FACT}
\FACT = \frac{|\boldsymbol{A'_{T}}|}{|\boldsymbol{A'_{\xi}}|}.      
\end{equation}

%%%%%%%%%%%%%%%%%%%%%%%%%%%%%%%%%%%%%%%%%%%%%%%%%%%%%%%%%%%%
\subsection{Numerical details}\label{sec:numerical}
%%%%%%%%%%%%%%%%%%%%%%%%%%%%%%%%%%%%%%%%%%%%%%%%%%%%%%%%%%%%%
The velocity field is represented by toroidal and poloidal scalar fields, which are expressed as radially varying Schmidt-normalised spherical harmonics. Radial variations are expressed using second-order finite differences on the zeros of Chebyshev polynomials. A predictor-corrector scheme is used for time stepping in spectral space that treats the diffusion terms implicitly. Further numerical details can be found in previous studies that use the same solver \citep{willis_2007,davies_2011,matsui_2016}. The solver has been validated against a benchmark solution proposed by \citet{breuer_2010} and a selection of simulations from \citet{tassin_2021}.

We have considered purely thermal, purely chemical, and thermochemical simulations. In most simulations, we have fixed $E=10^{-5}$, $\PrT=1$ and $Pr_\xi=10$ and sampled a region of the ($\sRaTinline$, $\sRaCinline$) plane that includes both thermally dominated and chemically dominated convection. Here, $\widetilde{Ra}_T^{c}$ and $\widetilde{Ra}_\xi^{c}$ are critical modified Rayleigh numbers at the onset of pure thermal and pure chemical convection, respectively, as reported in Table \ref{tab:RaC} in Appendix~\ref{app:Rac}.  The homogeneous purely thermal simulations reported in \citet{mound_2017} are also included (14 simulations, diamonds in Figure \ref{fig:force_regime}), as well as $7$ new purely compositional simulations (stars in Figure \ref{fig:force_regime}) with compositional Rayleigh varying from $\tRaC=10^{2}-10^{5}$ (right y-axis). The pure thermal ($\tRaC=0$) and chemical ($\tRaT=0$) simulations were plotted at $\sRaCinline=0.2$ and $\sRaTinline=0.2$, respectively, to enable the use of a log scale in Figure~\ref{fig:force_regime}. 

Furthermore, we report $49$ new homogeneous (i.e., $\qstar=0$) thermochemical models (triangles), with varying ($\tRaT,\tRaC$). For the thermal and chemical fields in these thermochemical simulations, we use the same Prandtl numbers as their pure convection counterpart (i.e., $\PrT=1$, $\PrC=10$, $\boldsymbol{Q}_{T,i}=\boldsymbol{Q}_{T,o}$ and $\boldsymbol{Q}_{\xi,o}=0$). In these simulations, $\tRaC$ is varied from $30-10^{5}$ at three fixed values of $\tRaT=90, 550\ \text{and}\ 1200$ ($39$ simulations, see top x-axis of Figure \ref{fig:force_regime}), while $\tRaT$ is varied from $30-4000$ for two fixed values of $\tRaC=300 \ \text{and}\  10^{4}$ ($10$ simulations).

We also investigate $26$ new heterogeneous thermochemical models with a spatial pattern of outer boundary heat flux inferred from seismic tomography \citep{masters_1996} where the degree of heterogeneity is fixed at $\qstar=5$.  We include $13$ pure thermal simulations from \citet{mound_2017}, which also use the same ``tomographic'' heat flux pattern with $\qstar=5$, with the other parameters same as the homogeneous models (i.e., $E=10^{-5}$, $Pr_T=1$ and $Pr_\xi=10$).

The strength and structure of thermal RILs are expected to be influenced by the amplitude of the heterogeneous boundary condition. Therefore, we also investigate simulations with $\qstar=\{2.3, 10\}$ in thermochemical simulations run at three values of thermal Rayleigh $\tRaT=\{90, 550, 1200\}$ and a fixed chemical Rayleigh $\tRaC=10000$ ($6$ simulations), and one simulation with $\qstar = 1$, $\tRaT = 1200$, $\tRaC = 10000$. 

Under conditions relevant to Earth’s core, chemical buoyancy is widely expected to dominate over thermal buoyancy, with estimates indicating that $\tRaC\gg\tRaT$ \citep{gubbins_2001,jones_2000}. Consistent with this expectation, the convective power available from compositional driving, which scales with $\GrCinline$, is inferred to be substantially larger than that associated with thermal driving, proportional to $\GrTinline$ \citep{lister_1995,davies_2015,Nimmo_2015}. Quantitative estimation of the Rayleigh number ratio for the bulk of the outer core may be complicated by the potential presence of a mushy layer at the inner core boundary \citep{wilczynski_2025}. Here, we focus on systematically exploring a region of the ($\tRaT,\tRaC$) parameter space to assess the differences between thermally and chemically dominated convection. Systematically investigating the effect of varying $E$ is beyond the scope of this study; however, we perform an initial check of Ekman dependence using $3$ additional thermochemical simulations with supercriticalities that target the rapidly rotating regime for $E=\{10^{-4}, 3\times10^{-5}, 10^{-6} \}$. 

All simulation parameters and diagnostics are tabulated in the supplementary table (Appendix \ref{app:diagnostics}). These simulation diagnostics, as defined in Appendix~\ref{app:SI} (section \ref{sec:diagnostics}), as well as the time-averaged results presented in section \ref{sec:results} are averaged over at least $100$ convective time-scales after the simulations reach a quasi-stationary state. 

%See ppt slide with a table of simulations \href{https://leeds365-my.sharepoint.com/:p:/r/personal/earsna_leeds_ac_uk/Documents/DD_project_files/tomographic/force_analysis_dd.pptx?d=w9da29ae239d644ba89f59c00824ab46c&csf=1&web=1&e=BmhK6i&nav=eyJzSWQiOjI2NywiY0lkIjoxOTkwNDgzNzA3fQ}{HERE}
 
%%%%%%%%%%%%%%%%%%%%%%%%%%%%%%%%%%%%%%%%%%%%%%%%%%%%%%%
\section{Results}\label{sec:results}
%%%%%%%%%%%%%%%%%%%%%%%%%%%%%%%%%%%%%%%%%%%%%%%%%%%%%%%
\begin{figure*}[h!]%% placement specifier
\centering
\includegraphics[width=0.77\textwidth]{\figfile{sims_E=1e-5_q=0}}
\caption{Regimes of force balance in homogeneous models. The symbol shapes indicate purely thermal (diamonds), purely chemical (stars) and thermochemical (left/right triangles). The symbols are coloured by the fluctuating curled force ratio $\CFIC$, indicating the role of inertia in the force balance relative to the Coriolis force. Open symbols are used for fluctuating force ratio $\FIC\geq0.1$, while filled symbols indicate $\FIC<0.1$. The thermochemical models are classified with the ratio of thermal to chemical buoyancy $\FACT \leq 1$ (left triangles) and $\FACT >1$ (right triangles), respectively. The dashed red lines at the lower left corner indicate the critical lines for the onset of convection. Blue and red squares indicate, respectively, the example chemically-dominated and thermally-dominated simulations discussed in the text.}\label{fig:force_regime}
\end{figure*}

Our main objective is to characterise the strength and thickness of the stable regions that emerge in our simulations. To scale these properties towards the extreme parameter regime that applies to Earth's core, we need to choose simulations with a geophysically relevant dynamical balance. In rotating convection, the turbulent flow regime with strong rotational constraints, known as the "rapidly-rotating" regime, is considered geophysically most relevant \citep{kunnen_2021}. We begin our analysis in section \ref{sec:force_regime} by studying the force balance to distinguish between various dynamical regimes of thermochemical convection. In particular, we investigate the role of non-linear inertia relative to the Coriolis force in the momentum equation (\ref{eqn:momentum_nd}). Our estimates of the relative importance of inertia $\FIC$ and $\CFIC$, as defined in section \ref{sec:force}, follow \citet{naskar_2025} to identify simulations that should be geophysically most relevant. The previous force analysis for pure thermal convection \citep{naskar_2025} has been extended to thermochemical convection in this study. In particular, we introduce a new metric to examine the relative importance of thermal vs. chemical buoyancy, $\FACT$, to determine which buoyancy source should decide the density distribution in the bulk of the core. We stress here that our thermal and chemical fields are different in two aspects: (a) the thermal field diffuses faster than the chemical field (i.e., $\PrT=1$ and $\PrC=10$) and (b) there are different boundary conditions for the two fields, with equal total heat flow across the ICB and CMB, whereas there is zero chemical flux across the CMB. Following the force analysis, in section \ref{sec:stable_region} we will present visualisations of two simulations with homogeneous boundary conditions: one simulation with $\FACT < 1$ (chemical buoyancy > thermal buoyancy), and one simulation with $\FACT > 1$ (thermal buoyancy > chemical buoyancy). These two simulations will then be compared with equivalent heterogeneous models with $\qstar=5$ that contain thermal RILs. Regional variations within the simulations are explored by comparing locally averaged radial profiles of $\NTsq$, $\NCsq$ and $\Nsq$ beneath selected reference locations (equatorial Africa, equatorial Pacific, equatorial Americas, North Pole) and globally averaged profiles. Finally, the strength and thickness of stratification are estimated from these profiles (section \ref{sec:strat_regime}) and the scaling of the RILs beneath Africa and the Pacific with model parameters is investigated in section \ref{sec:scaling}.    

%%%%%%%%%%%%%%%%%%%%%%%%%%%%%%%%%%%%%%%%%%%%%%%%%%%%%%%%%%%%%%
\subsection{Regimes of force balance}\label{sec:force_regime}
%%%%%%%%%%%%%%%%%%%%%%%%%%%%%%%%%%%%%%%%%%%%%%%%%%%%%%%%%%%%%%
Most of our simulations were run with $E=10^{-5}$, $Pr_T=1$ and $Pr_\xi=10$, varying the balance of chemical to thermal buoyancy and the strength of thermal boundary heterogeneity. We begin by considering the dynamics of simulations with homogeneous heat flux at the CMB ($\qstar=0$). The parameter regime of these simulations is shown in Figure \ref{fig:force_regime}, which also shows the role of inertia compared to the Coriolis force using two force ratio metrics $\FIC$ and $\CFIC$ (equation \ref{eqn:FIC}). The force ratio $\FIC$ measures the role of inertia compared to Coriolis force in the primary force balance, while the curled force ratio $\CFIC$ is an appropriate measure to investigate the relative role of inertia in the secondary balance.
The majority of these simulations are quasi-geostrophic (QG), having a primary balance between Coriolis and pressure terms, with a secondary contribution from inertia. For a few simulations at the highest $\tRaT$ considered, inertia enters the primary force balance (as measured by $\FIC>0.1$, equation \ref{eqn:FIC}) and the simulations are non-QG. As either $\tRaT$ or $\tRaC$ increases, the simulations gradually transition from non-turbulent to vigorously turbulent conditions as the relative importance of inertia increases compared to rotation in the secondary force balance (as measured by $\CFIC$). When the curled force ratio is of order one, convective motions are considered turbulent. Together, the two criteria help distinguish simulations that fall within the geostrophic turbulent flow regime—where the role of inertia is significant in the secondary balance (i.e., $\CFIC\sim \orderof{1}$), ensuring vigorous turbulence, but it does not enter the primary QG balance (i.e., $\FIC<0.1$), thereby maintaining rotational constraint \citep{naskar_2025}. The heterogeneous simulations cover a similar parameter space (see Appendix \ref{app:diagnostics}) and have similar force balances, although they have stronger thermal buoyancy at large length scales compared to their homogeneous counterparts.

%%%%%%%%%%%%%%%%%%%%%%%%%%%%%%%%%%%%%%%%%%%%%%%%%%%%%%%%%%%
\subsection{Stable regions}\label{sec:stable_region}
%%%%%%%%%%%%%%%%%%%%%%%%%%%%%%%%%%%%%%%%%%%%%%%%%%%%%%%%%%%
%Homogeneous sims
\begin{figure*}[h!]%% placement specifier
\centering
Chemically dominated $\color{Blue} \boldsymbol{(\square)}$\\
\vspace{5mm}
\begin{overpic}[width=0.32\textwidth,trim={1cm 2cm 2cm 2cm},clip]{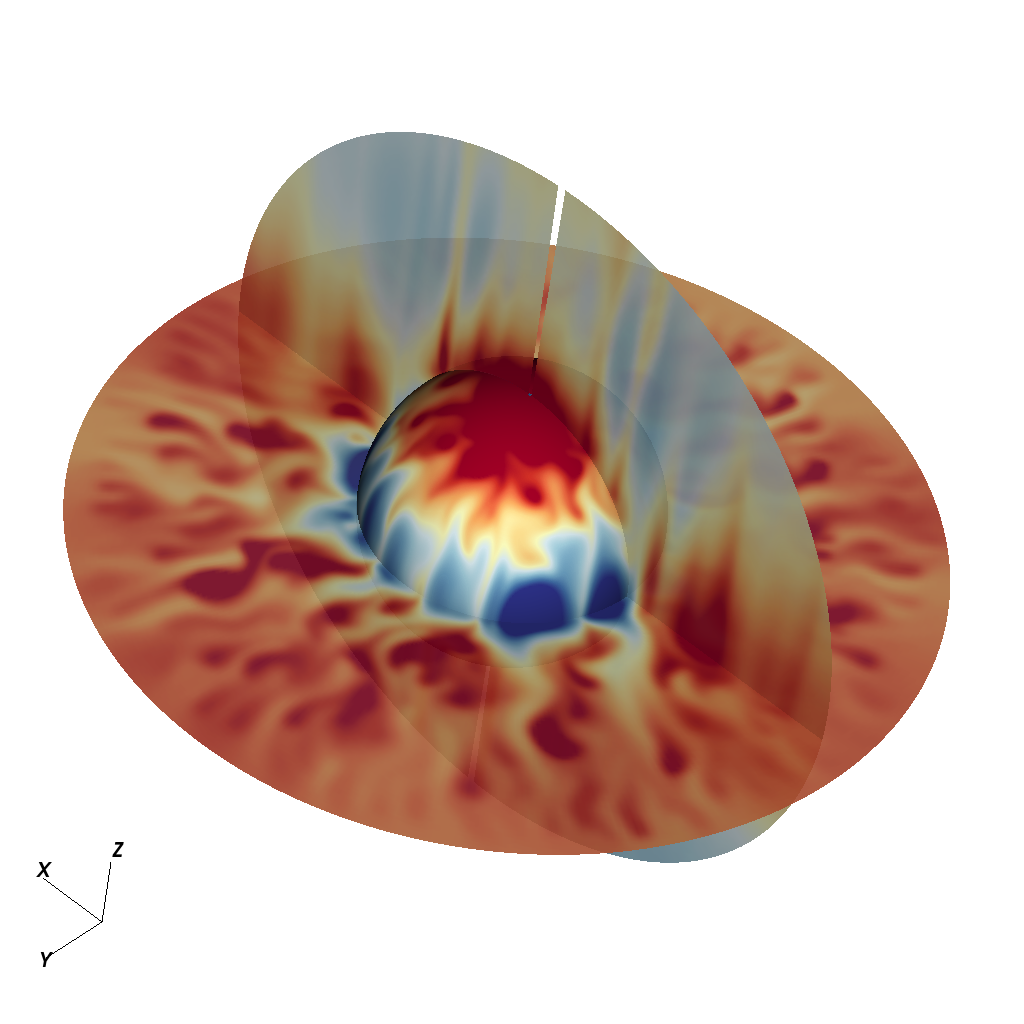}
\put(0,98){$(a)$}
\end{overpic} 
\begin{overpic}[width=0.32\linewidth,trim={1cm 2cm 2cm 2cm},clip]{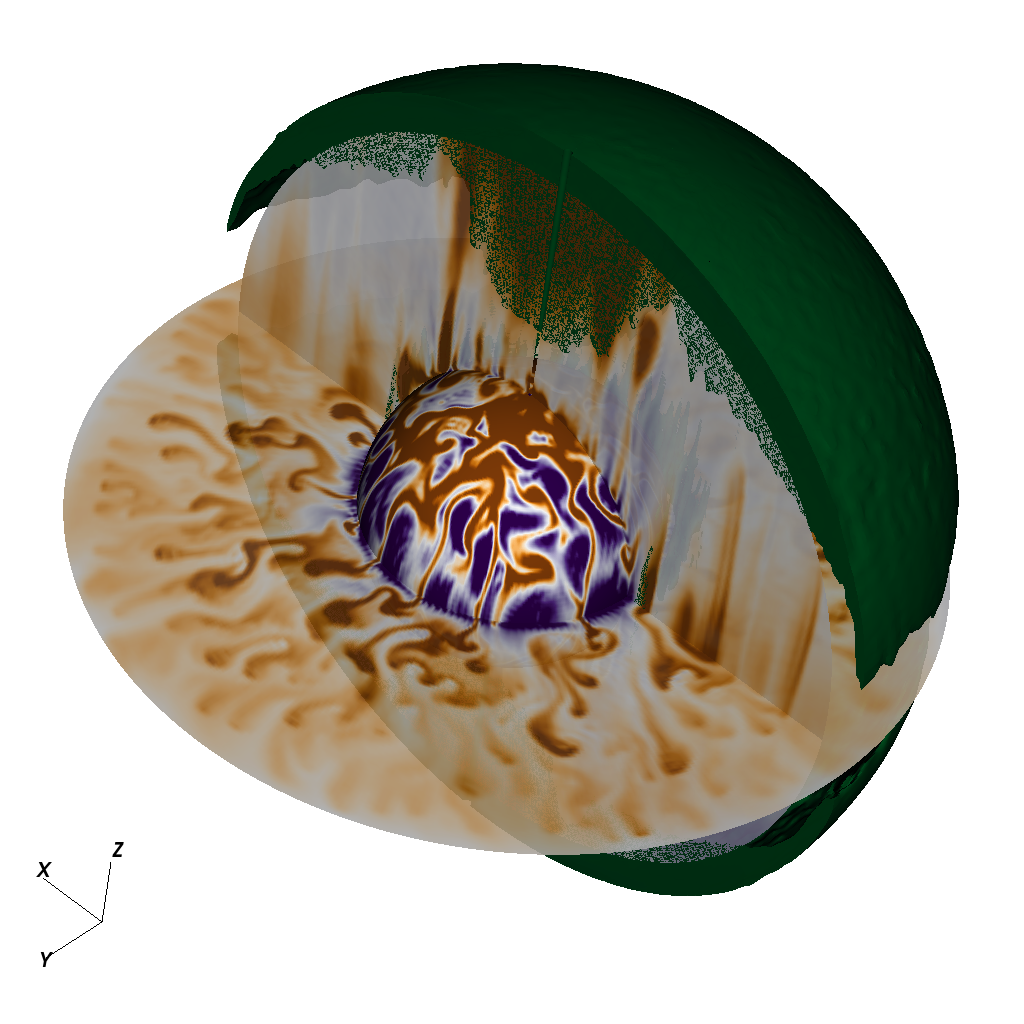}
\put(60,98){{\color{ForestGreen}$\Crd>0$}}
\put(0,98){$(b)$}
\end{overpic}
\begin{overpic}[width=0.32\linewidth,trim={1cm 2cm 2cm 2cm},clip]{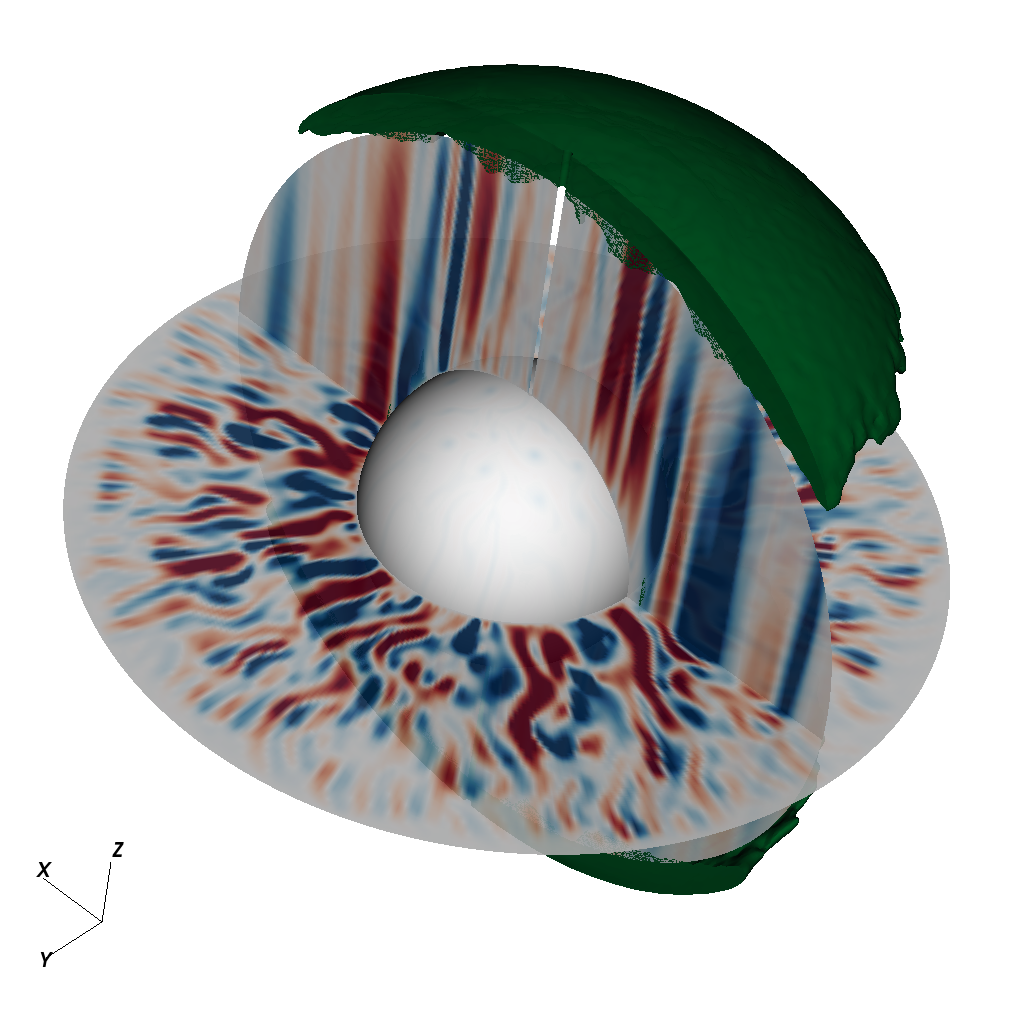}
\put(60,98){{\color{ForestGreen}$N^{2}>0$}}
\put(0,98){$(c)$}
\end{overpic}\\
%\begin{comment}
\vspace{1mm}
Thermally dominated $\color{BrickRed} \boldsymbol{(\square)}$\\
\vspace{5mm}
\begin{overpic}[width=0.32\linewidth,trim={1cm 2cm 2cm 2cm},clip]{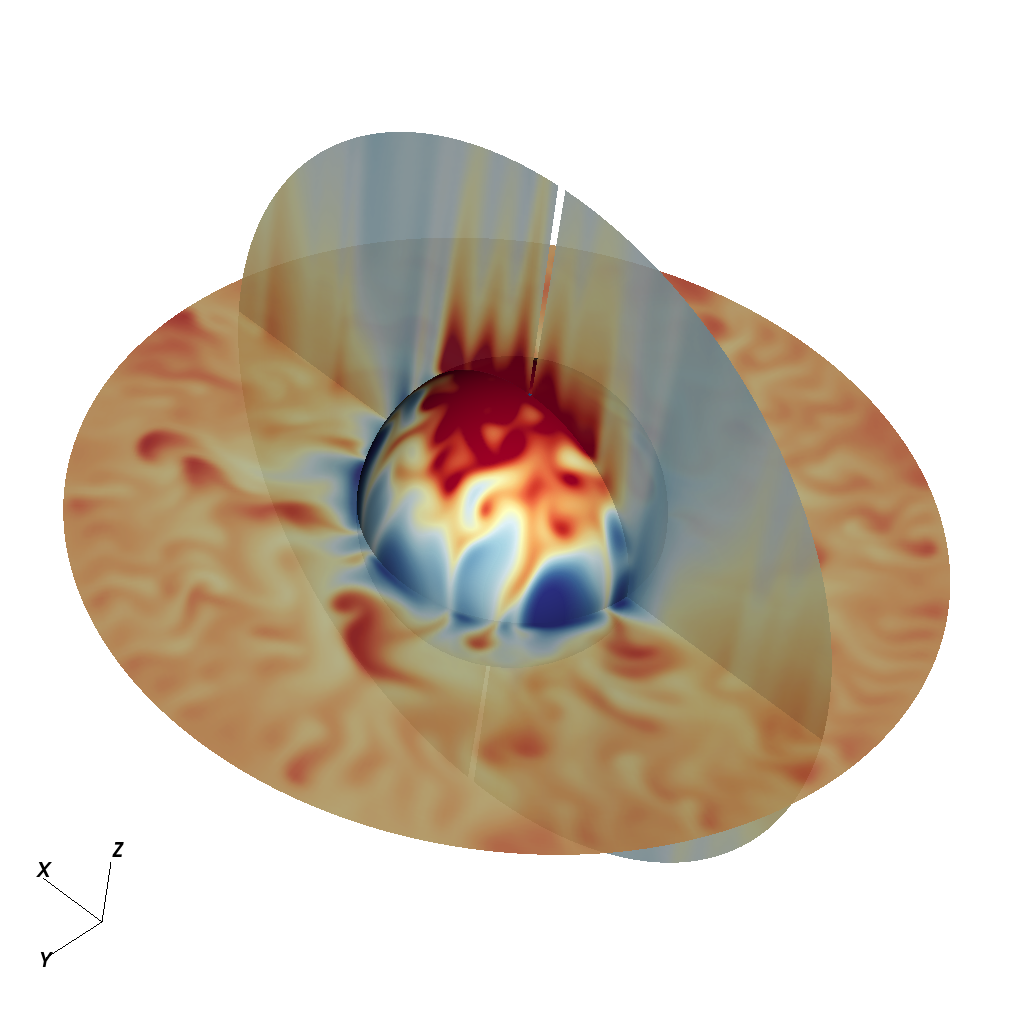}
\put(20,1){$\text{Thermal anomaly}$}
\put(0,98){$(d)$}
\end{overpic}
\begin{overpic}[width=0.32\linewidth,trim={1cm 2cm 2cm 2cm},clip]{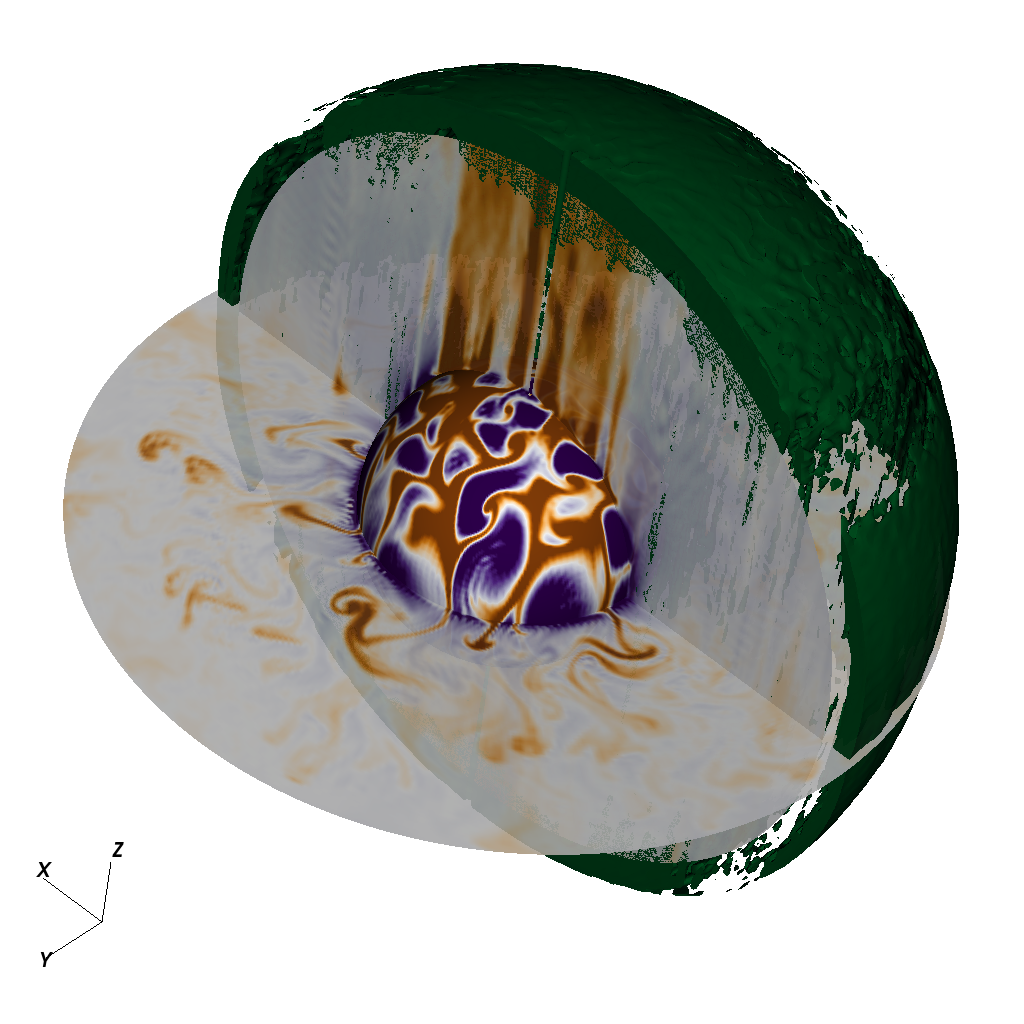}
\put(20,1){$\text{Chemical anomaly}$}
\put(0,98){$(e)$}
\end{overpic}
\begin{overpic}[width=0.32\linewidth,trim={1cm 2cm 2cm 2cm},clip]{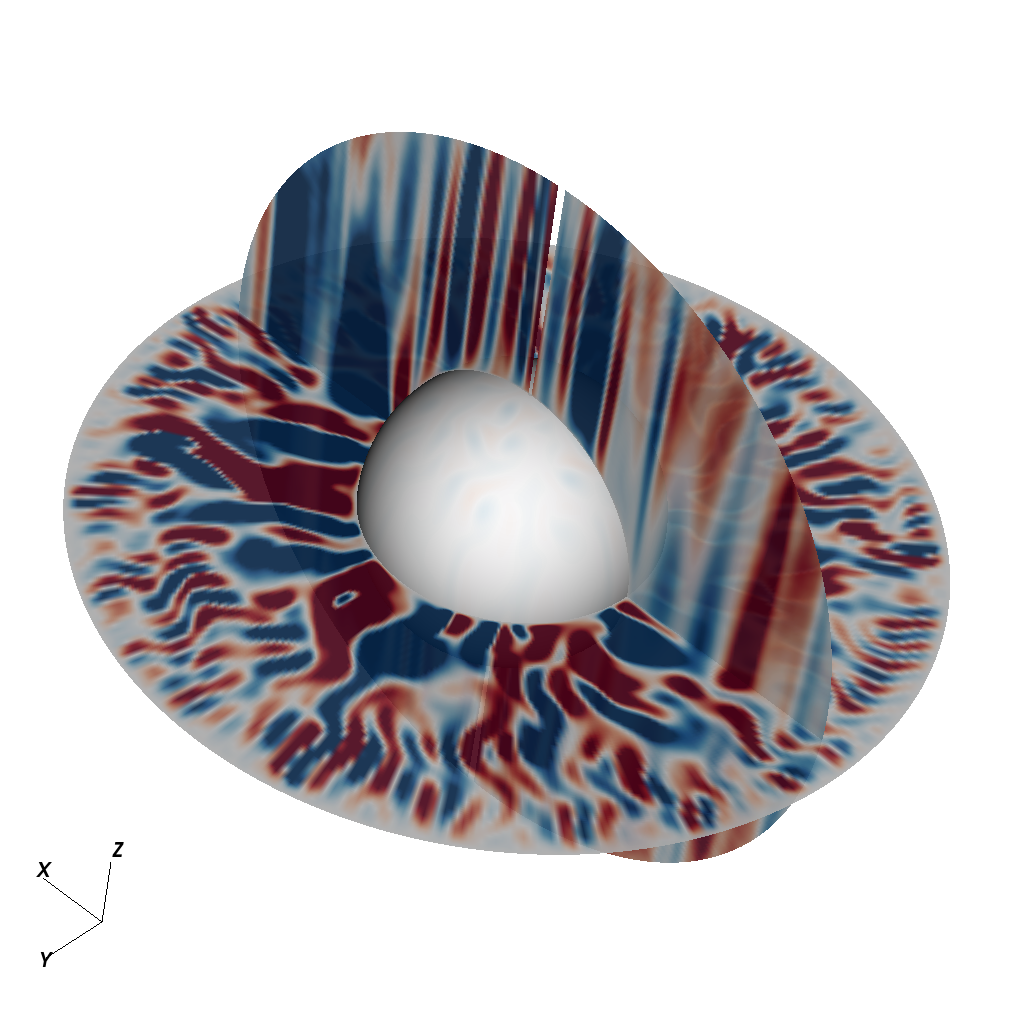}
\put(20,1){$\text{Radial velocity}$}
\put(0,98){$(f)$}
\end{overpic}
%\end{comment}
\caption{Scalar fields and their gradients in homogeneous simulations.  The equatorial, meridional and spherical ($r=0.54$) surfaces are colored by thermal anomaly (a,d), chemical anomaly (b,e), and radial velocity (c,f). The contour colours indicate hot (red) or cold (blue) fluid in (a,d), high (orange) or low (purple) concentration of light elements in (b,e), and positive (red) or negative (blue) radial velocity in (c,f), while grey represents zero velocity. The green iso-volume depicts time-averaged chemically stable regions with $\Crinline>0$ in (b,e), and convectively stable regions with $\Nsq>0$ in (c,f). Chemically dominated convection (a,b,c) at $\tRaT=90$ and $\tRaC=30000$, compared to thermally dominated convection (d,e,f) at $\tRaT=1200$ and $\tRaC=300$. To avoid obscuring the interior structure, the stable regions (green isovolume) are clipped in the front hemisphere.}\label{fig:homog}
\end{figure*}

\begin{figure*}[h!]%% placement specifier
\centering
Chemically dominated $\color{Blue} \boldsymbol{(\square)}$\\
\vspace{5mm}

\begin{overpic}[width=0.32\linewidth,trim={1cm 2cm 2cm 2cm},clip]{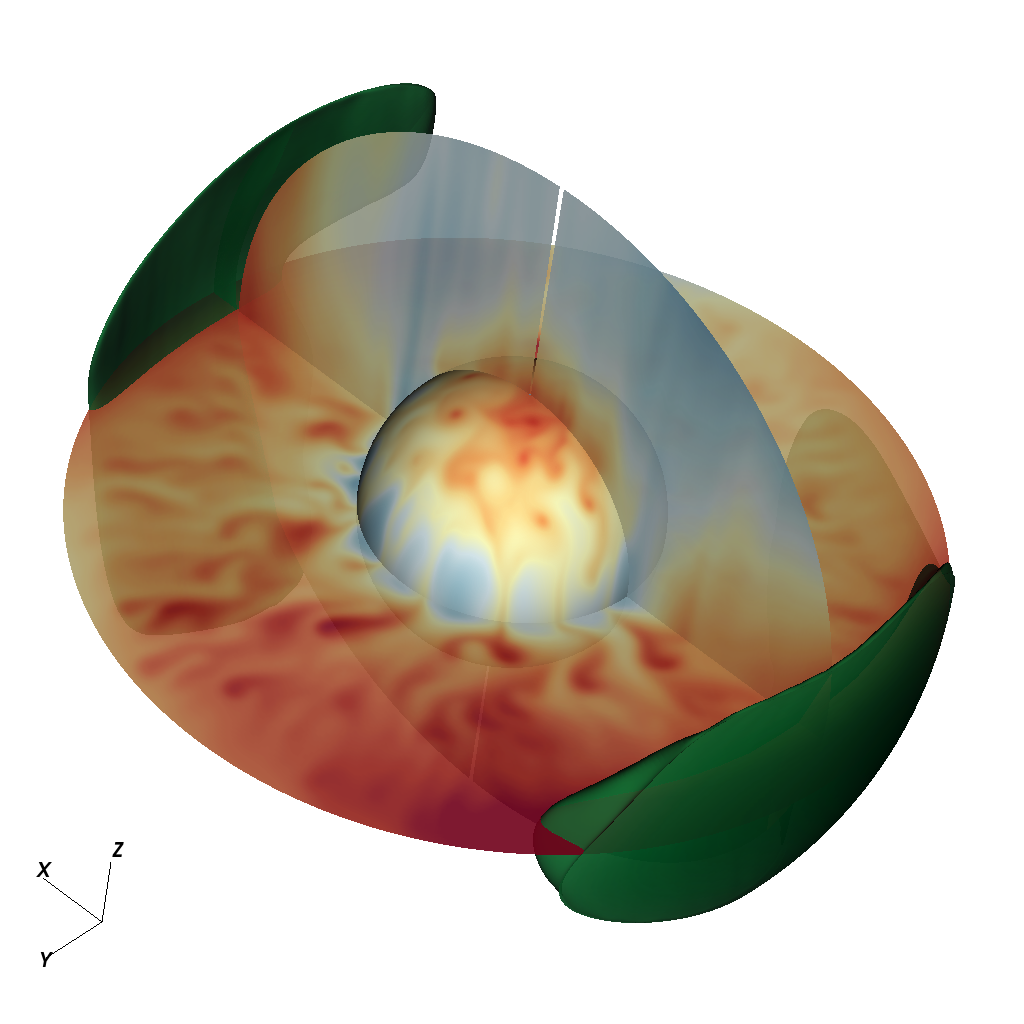}
\put(40,98){{\color{ForestGreen}$\Trd>0$}}
\put(0,98){$(a)$}
\end{overpic}%Newline after this is accounted for compilation 
\begin{overpic}[width=0.32\linewidth,trim={1cm 2cm 2cm 2cm},clip]{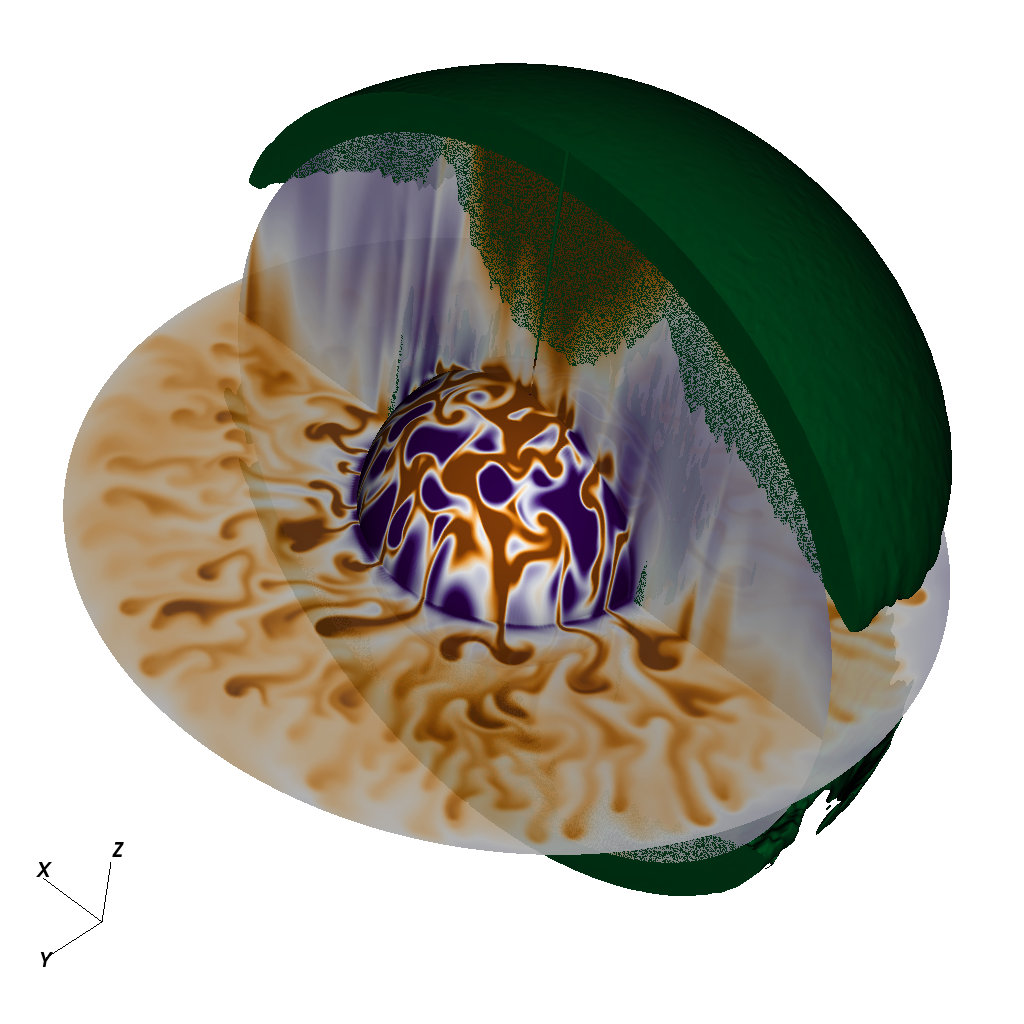}
\put(70,98){{\color{ForestGreen}$\Crd>0$}}
\put(0,98){$(b)$}
\end{overpic}
\begin{overpic}[width=0.32\linewidth,trim={1cm 2cm 2cm 2cm},clip]{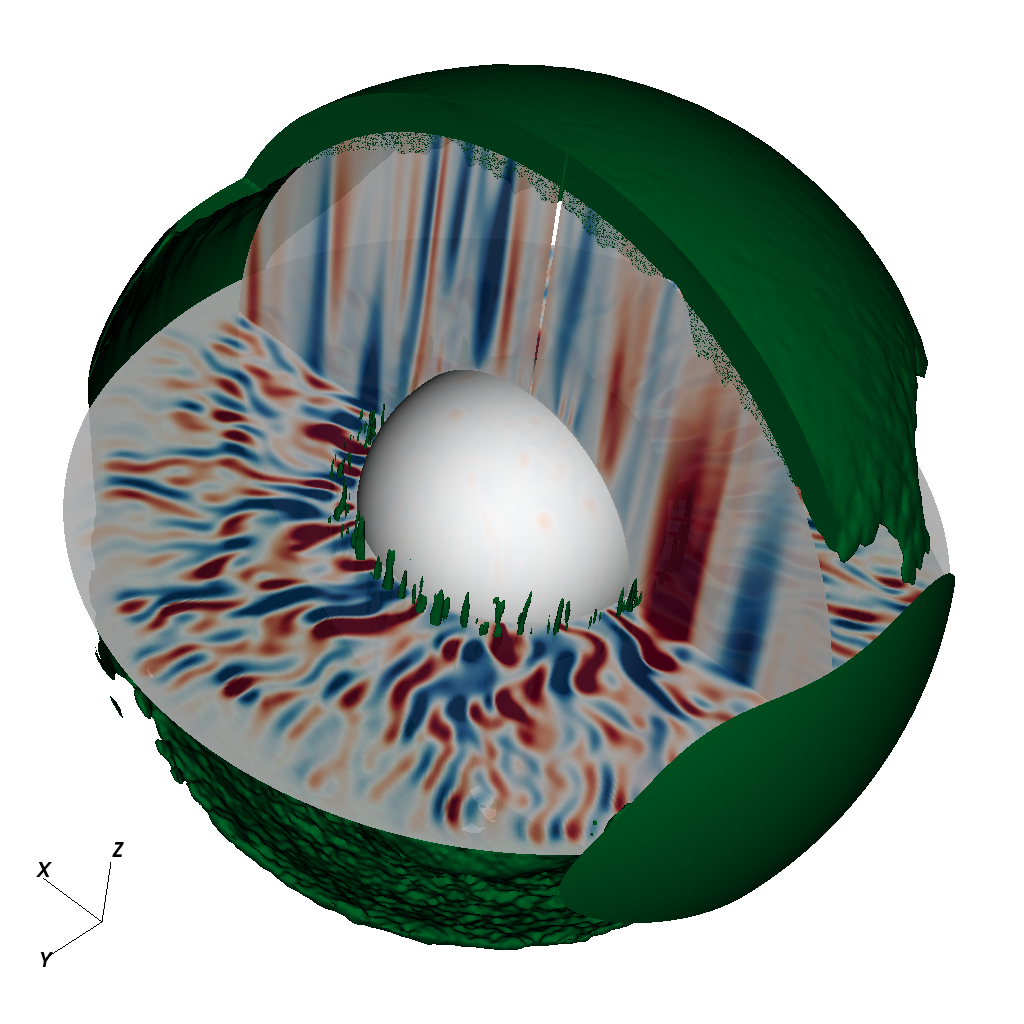}
\put(60,98){{\color{ForestGreen}$N^{2}>0$}}
\put(0,98){$(c)$}
\end{overpic}\\
\vspace{1mm}
Thermally dominated $\color{BrickRed} \boldsymbol{(\square)}$\\
\vspace{5mm}
\begin{overpic}[width=0.32\linewidth,trim={1cm 2cm 2cm 2cm},clip]{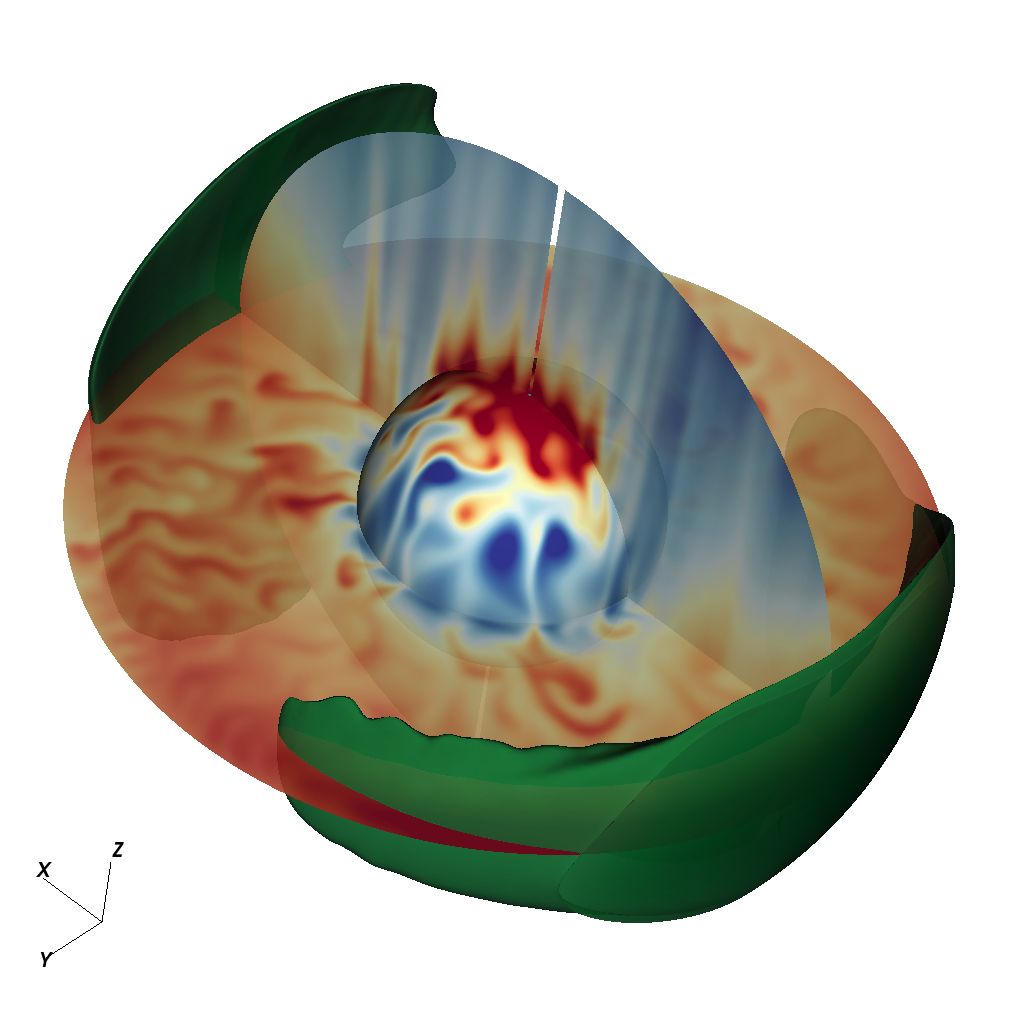}
\put(16,0){$\text{Thermal anomaly}$}
\put(0,98){$(d)$}
\end{overpic}
\begin{overpic}[width=0.32\linewidth,trim={1cm 2cm 2cm 2cm},clip]{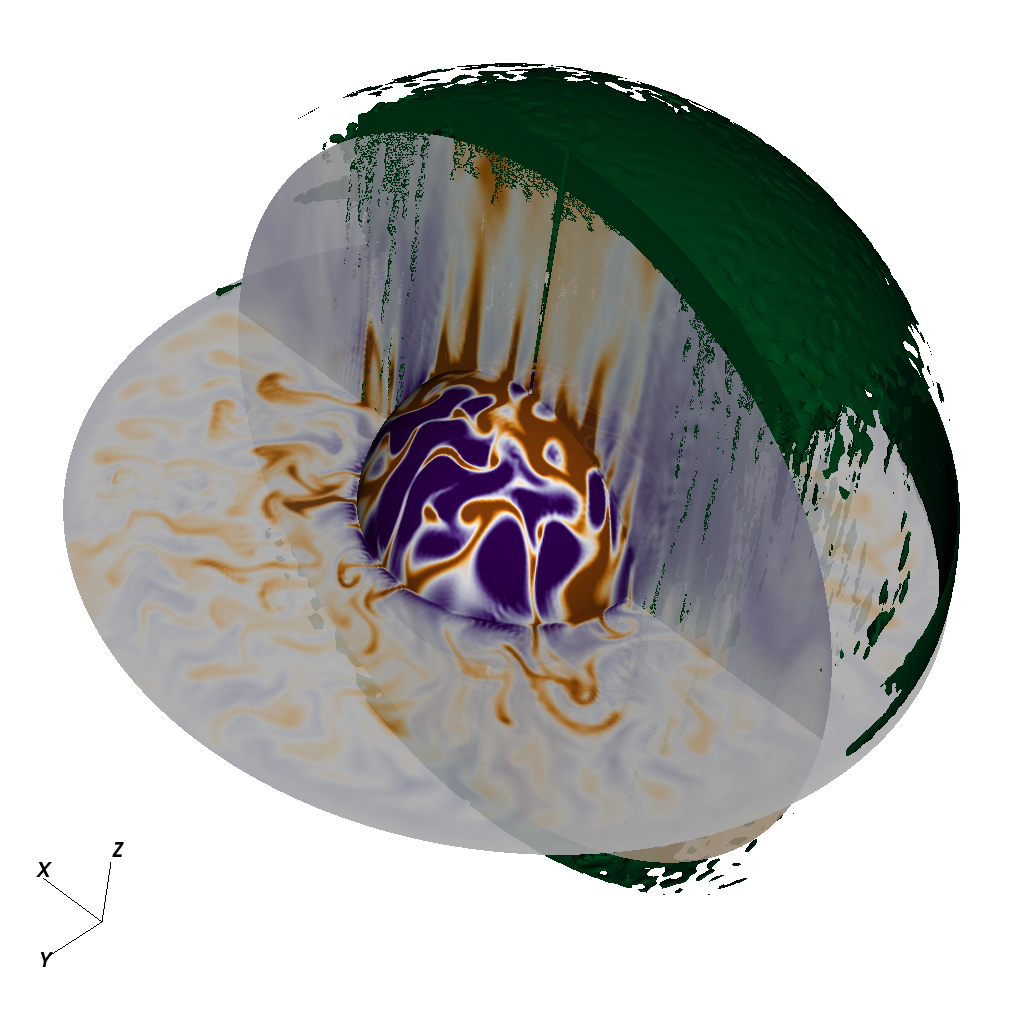}
\put(12,0){$\text{Chemical anomaly}$}
\put(0,98){$(e)$}
\end{overpic}
\begin{overpic}[width=0.32\linewidth,trim={1cm 2cm 2cm 2cm},clip]{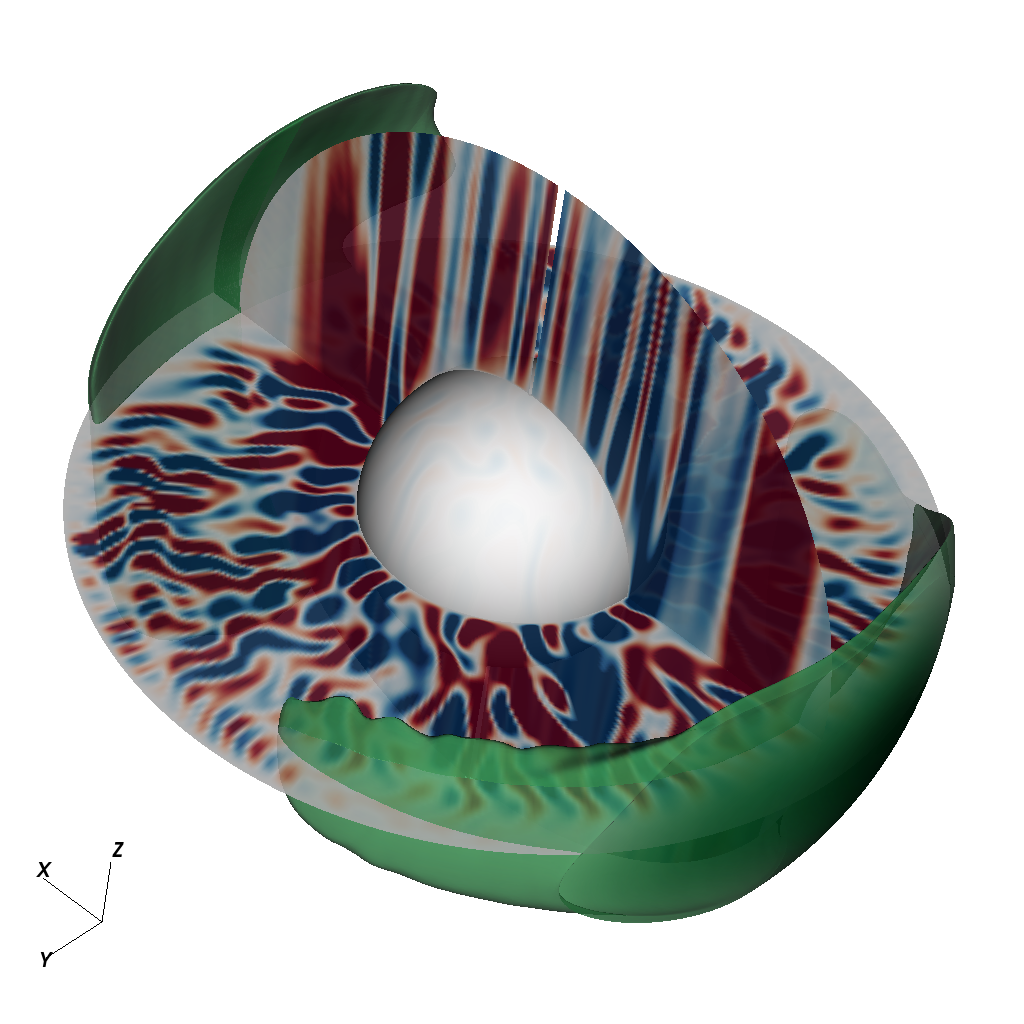}
\put(12,0){$\text{Radial velocity}$}
\put(0,98){$(f)$}
\end{overpic}\\
\caption{Scalar fields and their gradients in heterogeneous simulations. The interpretations of the contour colour and isovolumes are the same as in Figure~\ref{fig:homog}. These heterogeneous simulations for $\qstar=5$ have otherwise the same parameters as their homogeneous counterparts in Figure~\ref{fig:homog}. To avoid obscuring interior structures, the time-averaged stable regions (green isovolume) are clipped in the entire front hemisphere in the middle panels, and for $z>0.6$ in panel (c).}\label{fig:tomog}
\end{figure*}

With an overview of the global force balance established, the development of regions with inverted density gradients can now be investigated in various regimes of thermochemical convection. We begin with the visualisation of flow and scalar fields in four simulations: one chemically dominated and one thermally dominated, both with $\qstar=0$; and two corresponding simulations with $\qstar=5$, having otherwise the same parameters. These simulations are marked by red and blue symbols in Figure~\ref{fig:force_regime}, and instantaneous snapshots of thermal anomaly, chemical anomaly, and radial velocity fields are presented in Figures~\ref{fig:homog} and \ref{fig:tomog} for the homogeneous and heterogeneous simulations, respectively. In all cases, the thermal and chemical fields are correlated, as reported in previous studies \citep{trumper_2012,tassin_2021}, albeit with smaller length scales present in the chemical field due to the lower chemical diffusivity. Thermochemical plumes are visible in the equatorial slices, and the elongation of structures parallel to the rotation axis is evident in the meridional slices.

For all four of the cases in Figures~\ref{fig:homog} and \ref{fig:tomog}, there are regions near the CMB that, in the time average, have either a stabilising thermal or chemical gradient.  The total density gradient is a combination of thermal and chemical effects, and, depending on the simulation setup, both components can produce net stable regions; that is, regions with a positive value of the squared BV frequency $\Nsq>0$ (see equation \ref{eqn:Nsq_nd}).

%%%%%%%%%%%%%%%%%%%%%%%%%%%%%%%%%%%%%%%%%%%%%%%%%%%%%%%%%%%%%%%%%%%%%%
\subsubsection{Purely chemical stability}\label{sec:strat_chem}
%%%%%%%%%%%%%%%%%%%%%%%%%%%%%%%%%%%%%%%%%%%%%%%%%%%%%%%%%%%%%%%%%%%%

In the homogeneous simulations (Figure~\ref{fig:homog}), stable regions arise only due to chemical enrichment near the top of the core ($\Crinline > 0$). The extent of the chemically-enriched regions and the amplitude of $\Crinline$ tend to increase with $\tRaC$;
although they can form regardless of whether thermal or chemical driving of convection is more important (see also section~\ref{sec:scaling}). The regions with $\Crinline > 0 $ have maximum thickness and strength near the poles, similar to previous observations \citep{bouffard_2019}. A global layer with $\Crinline > 0 $ can form beneath the CMB in some cases; however, the chemically stratified regions can also be disrupted by sufficiently vigorous thermal convection. For our example homogeneous simulation with $\FACT < 1$, the region of overall convective stability ($\Nsq>0$) is more localised near the poles than the region with $\Crinline > 0$ (compare Figure~\ref{fig:homog}b and c). For the homogeneous simulation with $\FACT > 1$, the thermal driving overpowers the chemical effects, and there is no net convective stability (Figure~\ref{fig:homog}f), which is the case for most of our $\FACT > 1$ simulations at $\qstar=0$. 

The formation of chemically stable regions can be understood from the integrated balance of the terms in equation (\ref{eqn:composition_nd}), as presented in supplementary section \ref{sec:comp_bal}. Light element conservation requires that, once the solution reaches a steady state, the sum of the contributions from the convective flux, diffusive flux, and volumetric sink is independent of radius (see Appendix~\ref{app:SI} figure \ref{fig:comp_bal}). Therefore, the sum of convective and diffusive fluxes needs to match the energy flux unaccounted for by the sink ($\Delta\textrm{sink}$). For low supercriticalities, the diffusive flux remains a positive (i.e., outward) quantity near the topmost core. With increased forcing, the amplitude of advective transport increases and the interior fluid becomes increasingly well mixed, conductive flux in the fluid interior approaches zero, and the advective flux approaches $\Delta\textrm{sink}$.
With further increase in $Ra_\xi$, the advective flux locally rises above $\Delta\textrm{sink}$ and light element is delivered to the top of the core faster than it is assimilated.
In the simulations for which this occurs, excess light elements accumulate near the outer boundary, resulting in stable regions. These regions grow through time until they reach a steady state with the inverted chemical gradient ($\Crinline > 0$) driving the negative (i.e., inward) diffusive flux required for light element conservation.
The evolution of the system and the growth of the regions below the outer boundary to their steady-state thickness and level of enrichment occurs on the order of one viscous decay time. We analyse the layer properties in the steady state, noting that the saturation time of the layer properties should depend on the chemical Prandtl number, as well as the initial condition of the simulations.

%%%%%%%%%%%%%%%%%%%%%%%%%%%%%%%%%%%%%%%%%%%%%%%%%%%%%%%%%%%%%%%%%%%%%%
\subsubsection{Thermochemical stability}\label{sec:strat_regime}
%%%%%%%%%%%%%%%%%%%%%%%%%%%%%%%%%%%%%%%%%%%%%%%%%%%%%%%%%%%%%%%%%%%%

When the tomographic heat flux pattern with $\qstar = 5$ is applied to the outer boundary, regions with $\Trinline>0$ form at the top of the core beneath the African and Pacific LLVPs (Figure $\ref{fig:tomog}$, see also Figure~\ref{fig:geometry}). The morphology of these African and Pacific RILs is similar to that reported in previous pure thermal simulations \citep{mound_2019}. Radial flow near the top of the core exhibits considerable longitudinal variation in simulations with heterogeneous heat flux boundary conditions, for both $\FACT >1$ and $\FACT < 1$, due to the localised suppression of convection beneath the LLVPs. The zero chemical flux boundary condition tends to suppress convection everywhere near the top of the core, which is apparent in $\FACT < 1$ simulations. These observations can be more readily confirmed from the maps of radial velocity near the top of the core (Figure~\ref{fig:vel_radmap}). The morphology of stable regions and the accompanying pattern of flow near the top of the core depend on both the strength of the imposed thermal heterogeneity and the overall balance of thermal to chemical driving ($\FACT$).

%\subsubsection{Locally averaged radial profiles}\label{sub:strat_reg_prof}
\begin{figure*}[h!]%% placement specifier
\centering
Chemically dominated$\color{Blue} \boldsymbol{(\square)}$\hspace{5cm} Thermally dominated$\color{BrickRed} \boldsymbol{(\square)}$\\
\vspace{5mm}
\begin{overpic}[width=0.47\linewidth,trim={0cm 0cm 0cm 0cm},clip]{\figfile{Nsq_depth_Pm=0_Pr_T=1_Pr_C=10_q=0.0_E=1e-5_Ra_T=90_Ra_C=30000}} 
\put(0,78){$(a)$}
\end{overpic}
\begin{overpic}[width=0.47\linewidth,trim={0cm 0cm 0cm 0cm},clip]{\figfile{Nsq_depth_Pm=0_Pr_T=1_Pr_C=10_q=0.0_E=1e-5_Ra_T=1200_Ra_C=1000}}
\put(0,78){$(b)$}
\end{overpic}
\begin{overpic}[width=0.47\linewidth,trim={0cm 0cm 0cm 0cm},clip]{\figfile{Nsq_depth_Pm=0_Pr_T=1_Pr_C=10_q=5.0_E=1e-5_Ra_T=90_Ra_C=30000}}
\put(0,78){$(c)$}
\end{overpic}
\begin{overpic}[width=0.47\linewidth,trim={0cm 0cm 0cm 0cm},clip]{\figfile{Nsq_depth_Pm=0_Pr_T=1_Pr_C=10_q=5.0_E=1e-5_Ra_T=1200_Ra_C=1000}}
\put(0,78){$(d)$}
\end{overpic}
\caption{Variation of $\Nsq$ with depth below CMB near the top of the core for homogeneous (a,b) and heterogeneous (c,d) models, averaged around various locations (see Figure~\ref{fig:geometry} in Appendix~\ref{app:SI}) for models with compositionally dominated convection (a,c) at $\tRaT=90$ and $\tRaC=30000$, and thermally dominated convection (b,d) at $\tRaT=1200$ and $\tRaC=300$.}\label{fig:Nsq_prof}
\end{figure*}

Stable regions can arise due to both the thermal and chemical fields. Chemically stabilised regions preferentially form near the poles; however, they extend to low latitudes in all of our exemplar simulations (Figures $\ref{fig:homog}$ and $\ref{fig:tomog}$). Thermal RILs form below LLVPs when sufficiently strong heterogeneous boundary conditions are applied ($\qstar > 2$). Whether a local profile or the global average profile has a net density stratification (i.e., $\Nsq > 0$), depends on the combined effects of the thermal and chemical gradients. When $\FACT$ is greater than (less than) one, the thermal (chemical) gradient tends to set the net density profile (compare the locally averaged radial profiles of non-dimensional frequency $\Nsq$ of Figure \ref{fig:Nsq_prof} with the component profiles of Appendix~\ref{app:SI} Figures \ref{fig:homog_prof} and \ref{fig:tomog_prof}). The profiles of $\Nsq$ in the thermally dominated homogeneous simulation (Figure \ref{fig:Nsq_prof}b) have little dependence on location, and the chemical enrichment in the polar regions is too weak to greatly perturb the global average. In contrast, the polar regions in the chemically dominated homogeneous case (Figure \ref{fig:Nsq_prof}a) have a sufficiently large adverse gradient to produce a stable region in the globally averaged profile, even though the equatorial regions remain fully convecting. For the heterogeneous models (Figure \ref{fig:Nsq_prof}c,d), the profiles of $\Nsq$ beneath Africa and the Pacific indicate the existence of thermal RILs, and these are stronger in the simulation with $\FACT >1$. However, these thermal RILs are not sufficiently strong to make $\Nsq$ positive in the global average, although $\Nsq >0$ does occur in the heterogeneous simulation with $\FACT <1$ due to the polar chemical enrichment. In the simulations with $\FACT <1$ (e.g., Figure~\ref{fig:Nsq_prof}a,c), the local stable polar regions with $\Nsq > 0$ exist beneath a thin ($5-10\ \textrm{km}$) thermally convecting region, while the thermally stabilising gradients under equatorial Africa and Pacific extend to the CMB in the heterogeneous case. This LEA-driven stratification near the poles leaves a global signature underneath a convecting region ($15-20\ \textrm{km}$). The profiles of $\Nsq$ for the homogeneous and heterogeneous models differ primarily due to the presence of RILs beneath Africa and the Pacific. 

\begin{figure*}[h!]%% placement specifier
\centering
\begin{overpic}[width=0.7\linewidth,trim={0cm 0cm 0cm 0cm},clip]{\figfile{sims_E=1e-5_q=5_col_thk_noletter}}
\put(0,70){$(a)$}
\end{overpic}\\

\vspace{5mm}

\begin{overpic}[width=0.7\linewidth,trim={0cm 0cm 0cm 0cm},clip]{\figfile{sims_E=1e-5_q=5_col_str_noletter}}
\put(0,70){$(b)$}
\end{overpic}
\caption{Regimes of thermochemical stability in the heterogeneous models. The symbols are colored by their thickness (a) and strength (b). The thickness/strength in the colorbar represents the maximum value among the locally averaged profiles we considered (e.g see Figure~\ref{fig:tomog_prof}). The globally averaged profile exhibits stratification in the model represented by a circle, though some regions (e.g., beneath America) remain fully convecting (Figure \ref{fig:Nsq_prof}c). A complete global layer of stratification is found for the model indicated by a diamond symbol.}\label{fig:regime_strat_q5}
\end{figure*}

For most heterogeneous models ($\qstar=5$), the thermal RILs beneath the Pacific and Africa are the thickest (Figure~\ref{fig:regime_strat_q5}a) and strongest (Figure~\ref{fig:regime_strat_q5}b) stable regions among the six locations considered for local averaging (see section \ref{sec:diagnostics}). However, for three cases with $\{\tRaT,\tRaC\}\in\left(\{30,10000\},\{90,30000\},\{90,100000\}\right)$ and $\FACT <1$, the most strongly stabilised regions are due to LEA near the poles. With increasing chemical forcing ($\tRaC$), the region of LEA-driven stability increases in thickness, strength, and extent. These chemically stabilised regions can be sufficiently strong to make $\Nsq>0$ in the globally averaged profile, and with further increase in $\tRaC$, a global stable layer develops. Notably, the stable regions of both thermal and chemical origin can be $\orderof{100}$ km thick with normalised BV frequency, $\N\sim\orderof{0.1}$. The thermal RILs are the dominant stable regions for most of our simulations, and their strength is only weakly sensitive to $\tRaC$ in the investigated parameter regime. This can be attributed to our choice of a chemically neutral boundary condition at the top of the core (i.e., $\Crinline=0$ at $r=r_o$), in contrast to the thermal field that is either fully convecting over the whole CMB (for $\qstar=0$ ) or only locally stable (for $\qstar=5$). Therefore, convective variations near the top of the core are not as strongly impacted by the chemical boundary condition. Consequently, we will mainly focus on the properties of the African and Pacific RILs as a function of $\tRaT$, noting that the strength of chemical forcing can also affect their properties. 

%%%%%%%%%%%%%%%%%%%%%%%%%%%%%%%%%%%%%%%%%%%%%%%%%%%%%%%%%%%%%%%%%%%%%%
\subsubsection{Scaling of RIL properties}\label{sec:scaling}
%%%%%%%%%%%%%%%%%%%%%%%%%%%%%%%%%%%%%%%%%%%%%%%%%%%%%%%%%%%%%%%%%%%%%%
\begin{figure*}[h!]%% placement specifier
%\centering
%\qquad
%(a)
\hspace{1mm}
\begin{overpic}[width=0.47\linewidth,trim={0cm 0cm 0cm 0cm},clip]{\figfile{strat_thk_scaling_Africa_avg_Pm=0_Pr_T=1_Pr_C=10_q=5.0_E=1e-5}}
\put(0,87){$(a)$}
\end{overpic}
%(b)
%\hspace{2mm}
\begin{overpic}[width=0.51\linewidth,trim={0cm 0cm 0cm 0cm},clip]{\figfile{strat_thk_scaling_Pacific_avg_Pm=0_Pr_T=1_Pr_C=10_q=5.0_E=1e-5}}
\put(-2,80){$(b)$}
\end{overpic}\\
%
%(c)
\vspace{2mm}
\begin{overpic}[width=0.47\linewidth,trim={0cm 0cm 0cm 0cm},clip]{\figfile{strat_str_scaling_Africa_avg_Pm=0_Pr_T=1_Pr_C=10_q=5.0_E=1e-5}}
\put(0,83){$(c)$}
\end{overpic}
%
%(d)
%\hspace{2mm}
\begin{overpic}[width=0.47\linewidth,trim={0cm 0cm 0cm 0cm},clip]{\figfile{strat_str_scaling_Pacific_avg_Pm=0_Pr_T=1_Pr_C=10_q=5.0_E=1e-5}}
\put(-2,83){$(d)$}
\end{overpic}
\caption{Scaling of thickness (a,b) and strength (c,d) of the stable regions averaged around Africa (a,c) and the Pacific (b,d). The symbol shapes indicate purely thermal (diamonds) and thermochemical (left/right triangles) simulations. The thermochemical models are classified with the ratio of thermal to
chemical buoyancy $\FACT < 1$ (left triangles) and $\FACT > 1$ (right triangles), respectively. The simulations with $\FIC>0.1$ or $\sRaTinline<8$, marked by open symbols, have been excluded from the fit \citep{long_2020,naskar_2025}.}\label{fig:scaling}
\end{figure*}

The thickness and strength of the African and Pacific RILs generally increase with $\tRaT$ (Figure~\ref{fig:scaling}). However, the trend also depends on the ratio of thermal to chemical buoyancy forcing, especially when considering the thickness of the RILs. For $200<\tRaT<3000$, RIL thickness approximately increases as $\delta\sim\tRaT^{0.2}$ and RIL strength closely follows a $\N\sim\tRaT^{0.5}$ scaling, which can be compared to previous theoretical scaling laws proposed for purely thermal convection \citep{mound_2020}. In the construction of these empirical fits, simulations with $\tRaT<200$ have been excluded as either (a) the models are too weakly turbulent, following the criteria presented in \citep{long_2020}, or (b) the strongly turbulent models with high $\tRaC$ have LEA-driven globally stable regions, making the properties incomparable to the theoretical scaling laws proposed for purely thermal RILs. We also exclude simulations with high thermal forcing ($\tRaT>3000$) as they are non-QG ($\FIC>0.1$), according to the criteria described in equation \ref{eqn:FIC}. Despite the exclusion of simulations with strong chemical stabilisation, the empirical scaling exponent of $0.2$ for the thickness of these thermochemical RILs is approximately double the theoretical prediction (i.e., $\delta\sim\tRaT^{0.1}$ \citep{mound_2020}). The strength of the RILs almost perfectly follows the theoretical estimate (i.e., $\N\sim\tRaT^{0.5}$), except for one model with global chemical stability (diamond in Figure \ref{fig:regime_strat_q5}), for which $\N$ is approximately $3$ times higher than other simulations with the same value of $\tRaT$. The RIL strength and thickness also increase with increasing $\qstar$, though more simulations at varying $\qstar$ should be performed to establish reliable scaling relations.

%%%%%%%%%%%%%%%%%%%%%%%%%%%%%%%%%%%%%%%%%%%%%%%%%%%%%%%%%%%%
\section{Discussion and conclusions}\label{sec:conclusions}
%%%%%%%%%%%%%%%%%%%%%%%%%%%%%%%%%%%%%%%%%%%%%%%%%%%%%%%%%%%%
\begin{figure*}[h!]%% placement specifier
\centering
\begin{overpic}[width=0.32\linewidth,trim={1cm 2cm 2cm 2cm},clip]{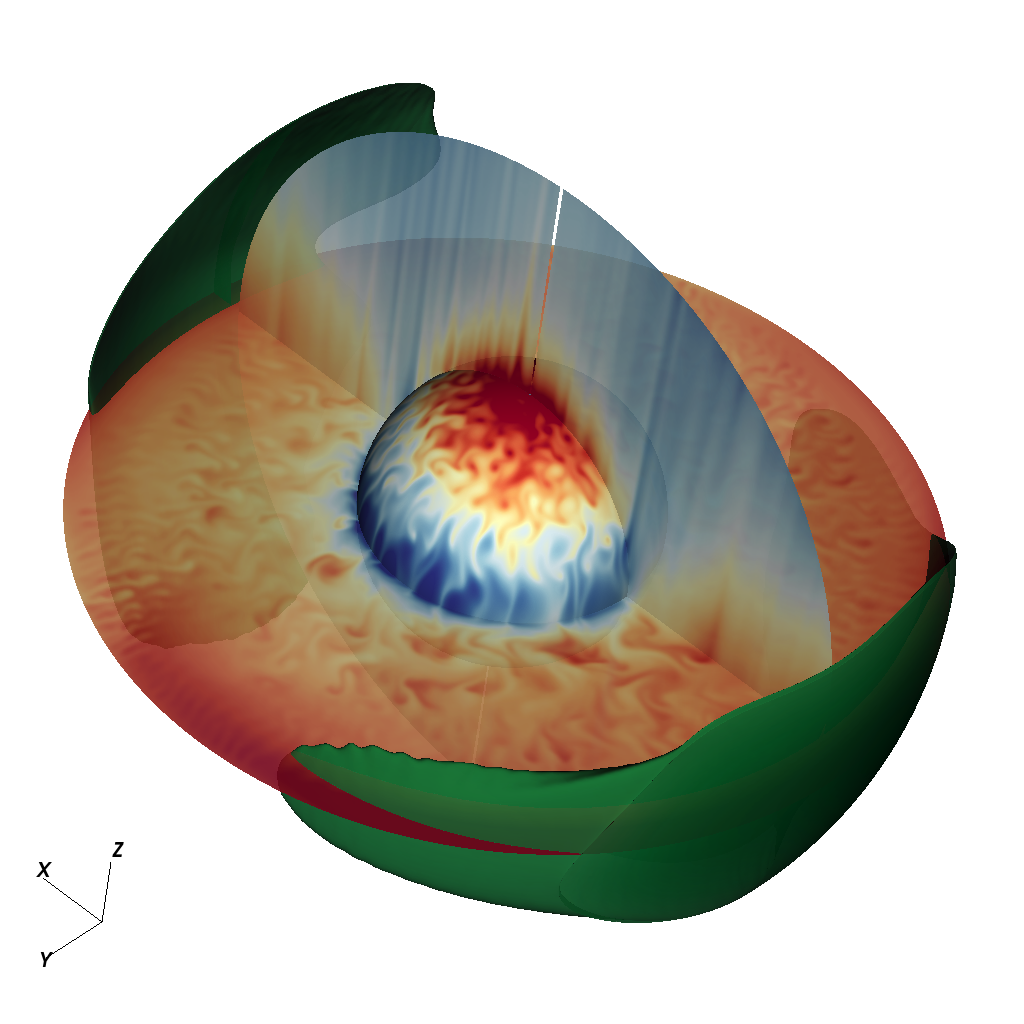}
\put(0,90){$(a)$}
\end{overpic}
\begin{overpic}[width=0.32\linewidth,trim={1cm 2cm 2cm 2cm},clip]{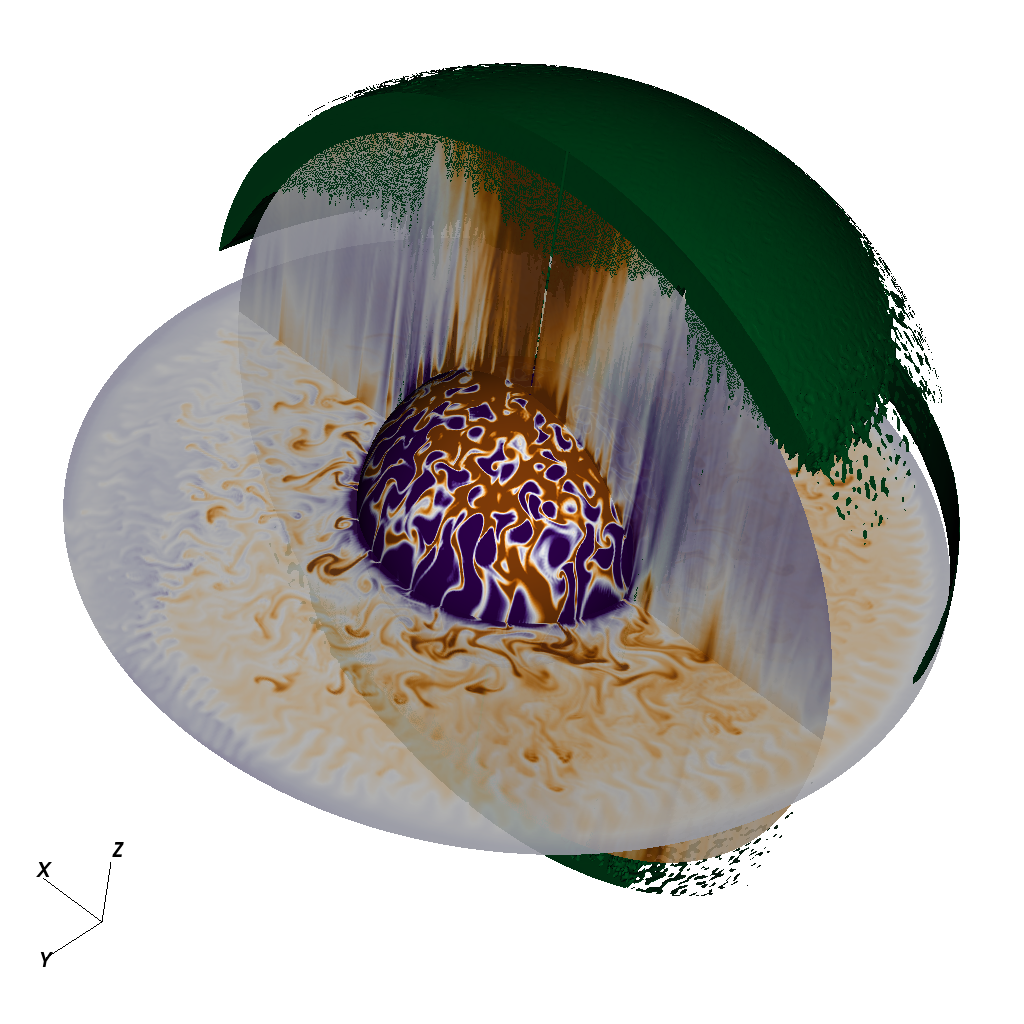}
\put(0,90){$(b)$}
\end{overpic}
\begin{overpic}[width=0.32\linewidth,trim={1cm 2cm 2cm 2cm},clip]{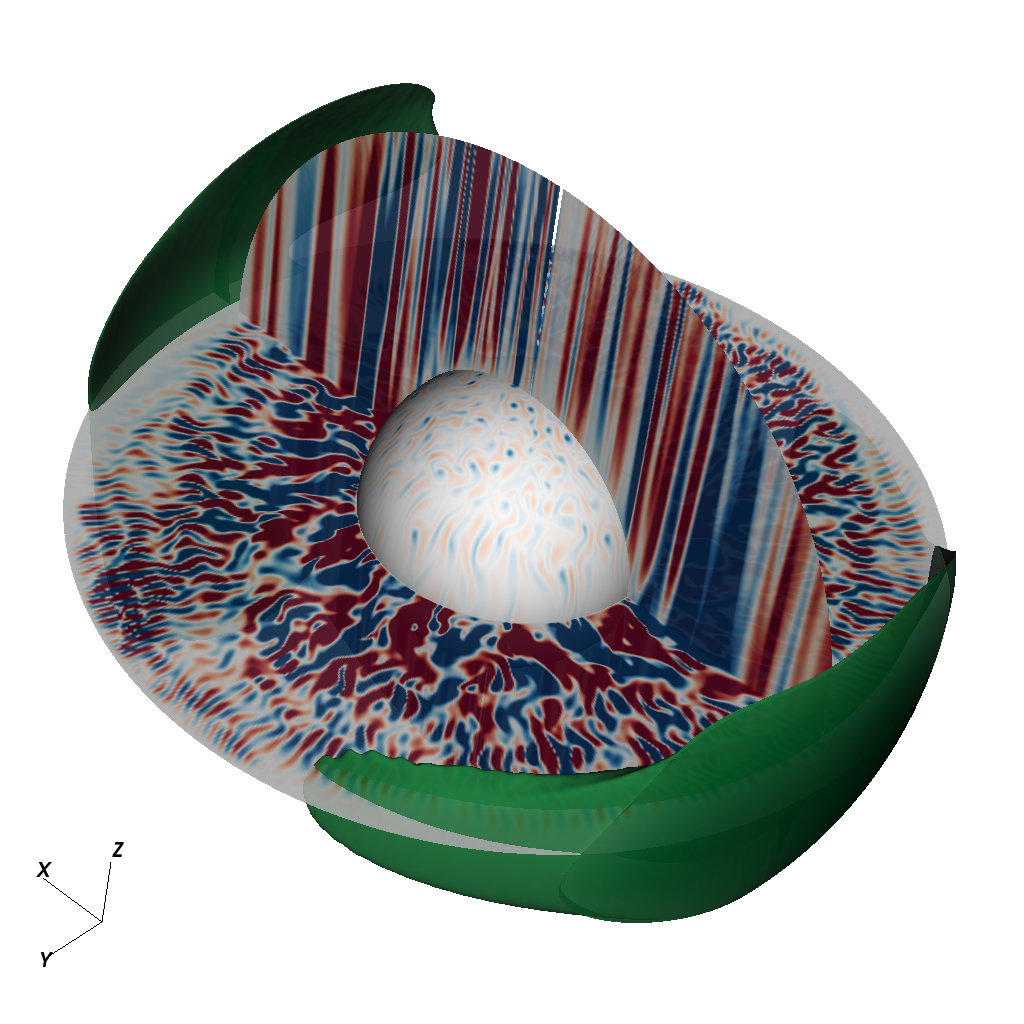}
\put(0,90){$(c)$}
\end{overpic}
\caption{Scalar fields and their gradients in a heterogeneous model at $E=10^{-6}$.  Contours and isovolumes have the same meaning as in Figure~\ref{fig:tomog}. The Rayleigh numbers for this model are $\tRaT=2000$ and $\tRaC=30000$. The time-averaged stable regions (green isovolume) are clipped in the front hemisphere in panel (b) for better visualisation of the internal dynamics.}\label{fig:tomog_e6}
\end{figure*}

We have used numerical simulations of thermochemically-driven convection in a rotating spherical shell to investigate the formation and properties of thermally and/or chemically stable regions at the top of Earth's outer core. Unlike previous studies where thermal and chemical buoyancy are treated together as "codensity", here we have imposed a laterally varying heat flux boundary condition derived from seismic tomography on the temperature field alone, while zero chemical flux is assumed at the CMB. This geophysically relevant setup allows thermally and chemically stable regions to form and evolve separately, and also allows the destabilising buoyancy from one component to influence the stabilising regions induced by the other component. Our main finding is that thermally stable Regional Inversion Lenses (RILs), which were previously identified in simulations of thermal convection \citep{mound_2019}, are also a robust feature of our thermochemical simulations. These RILs form at low latitudes under the Pacific and Atlantic Large Low Velocity Provinces, where the boundary heat flow is anomalously low. 

In our simulations, RILs are $\orderof{100}$~km-thick with a normalised BV frequency of $\Nsq \sim\ 0.1$ similar to our previous work \citep{mound_2019,mound_2020} and are characterised by a stabilising overall density gradient (driven by the thermal field) and weak upwelling flow. We emphasise here that the radial flow near the outer boundary is not completely suppressed inside the stratified regions in any of our simulations (e.g., see Figure \ref{fig:vel_radmap}). The radial penetration of the flow inside the RILs depends on their stratification strength $\Nsq$, and scaling this property to core conditions to infer radial motion near the top can be difficult, as discussed later in this section. The boundary heterogeneity also drives zonal flows, with velocity magnitudes similar to the deep jets associated with columnar convection in our homogeneous simulations. This similarity is evident in the comparable zonal velocity amplitudes for both homogeneous and heterogeneous models (Figure \ref{fig:zonal}), which broadly follow an established scaling law, relating the zonal velocity with input convective power, based on a balance between Reynolds stresses and Ekman friction \citep[for details see section \ref{sec:zonal} and][]{aubert_2005}. Considering uncertainties (e.g., resulting from dependence on $\tRaC$, boundary conditions, large Ekman number), the best fit scaling exponent $0.6$ is reasonably close to the theoretical scaling exponent $0.8$. In the presence of a magnetic field, the zonal flow is expected to follow a shallower scaling with exponent $0.5$ that gives a reasonable extrapolated estimate for zonal flow amplitude at the top of Earth's core \citep{aubert_2005}.    

Our simulations show thermochemically stable regions with varying characteristics. Chemically driven local or global stable regions arise in our homogeneous simulations when chemical buoyancy dominates thermal buoyancy. In this parameter regime, these regions co-exist with thermally driven RILs in our heterogeneous models. The strength and thickness of the chemically stable regions increase with the compositional Rayleigh number and can reach values comparable to those for the thermal RILs in the parameter space studied. The strength and thickness of the thermally stabilised regions increase with the thermal Rayleigh number. The thermal RIL strength scales as $\tRaT^{0.5}$, similar to pure thermal models \citep{mound_2020}; however, the RIL thickness follows a steeper ($\sim\tRaT^{0.2}$) scaling compared to the theoretical $\tRaT^{0.1}$ scaling, the difference being attributed to the chemical driving. 

Chemically stable regions in our simulations arise naturally from the internal dynamics of the rotating convection rather than through an imposed external mechanism, as is often assumed \citep[e.g.,][]{buffett2010stratification, landeau_2016, brodholt_2017}. \citet{bouffard_2019} found global dynamically-induced chemically stable layers in simulations of pure chemical rotating spherical shell convection. These layers were non-stationary and grew slowly on the viscous diffusion timescale. By contrast, the chemically stable regions in our simulations reach a steady state in about a viscous diffusion timescale. In the steady state, the downward diffusive chemical flux (together with some downward entrainment) balances upward advection of light material into the stable region. \citet{bouffard_2019} implemented a novel numerical method to neglect chemical diffusion entirely in order to approximate the high $\PrC$ limit appropriate to Earth's core and suggest that diffusion should remain non-negligible even at $\PrC\sim1000$. Although this is larger than the value $\PrC=10$ adopted in our simulations to reduce computational costs, the role of the downward diffusive chemical flux in the balance should remain important at higher $\PrC$.  The chemically stable regions in both studies exhibit latitudinal variations with maximum thickness and strength near the poles, which gradually reduce towards the equator. In our simulations, this effect seems to be caused by enhanced vertical transport of chemical plumes inside the tangent cylinder (Figure \ref{fig:comp_flux}) combined with meridional circulation, which transports light elements away from the poles. Additionally, higher stirring and mixing outside the tangent cylinder than inside, and/or thermal convection eroding the layer at the equator may also play a role \citep{Bouffard2020ConvectiveCore}.

Both the fingering (thermally stable but compositionally unstable RILs below Africa and Pacific) and semi-convection (thermally unstable but compositionally stable RILs at the poles) regimes of double-diffusive convection might coexist in our simulations for at least some parameters. In the fingering regime \citep{guervilly_2022,tassin_2024}, the resulting flow structures depend on the relative strength of stratification and rotation, represented by $\Nsq$. Our simulations should fall in the rotation-dominated columnar convection regime, and therefore, gravity-aligned finger structures are unlikely to develop in the equatorial RILs. Similarly, although turbulent semi-convection \citep{pruzina_2025} may produce a stable layer overlaying a deeper convecting region, direct comparisons with these studies are difficult due to differences in setup and the local nature of the stratification in our study. Flow structures are columnar in both the potential fingering and semi-convection regions of our simulations, indicating the dominant effect of rotation, and we do not observe any obvious signature of either double-diffusive dynamic structure inside the stratified regions.

Our simulations are set up to mimic a geophysically realistic distribution of thermochemical buoyancy and access a regime of rapidly rotating geostrophic turbulence that is broadly characteristic of Earth's core. However, they lack a magnetic field and did not reach the high value of $\PrC$ and the low value of $\PrT$ that characterise Earth's core dynamics. Our previous thermally-driven dynamo simulations showed that strong and thick RILs were preserved from corresponding non-magnetic cases \citep{mound_2023}, which supports theoretical arguments that the magnetic field does not dictate the gross properties of the RILs \citep{mound_2020}. The results of \citet{bouffard_2019} suggest that thick and strongly stable chemical regions at the top of the core emerge with increasing $\PrC$, which suggests that the stable regions found in our simulations would not be destroyed in the limit $\PrC \rightarrow \infty$. The strength of the chemically stable region near the poles generally increases with increasing $\tRaC$ and decreases with increasing $\tRaT$, since sufficiently strong thermal driving can disrupt such regions (Figure \ref{fig:LEA_scaling}). The $E$-dependence of the solutions is hard to assess because the geophysically relevant value of $E=10^{-15}$ is not computationally accessible through direct numerical simulation. Our one, thermally dominated, simulation at $E=10^{-6}$ (Figure~\ref{fig:tomog_e6}) has stable regions that are morphologically similar to counterparts at higher Ekman number, with a thickness and strength of $\delta_{max}\sim213$~km and $\N\sim0.03$. No simple trend in these properties was found when comparing this simulation to counterparts with $E=10^{-5}$, $3\times10^{-5}$ and $10^{-4}$ (Figure~\ref{fig:Ek_dep}). The thickness and strength of RILs also exhibit a broadly increasing trend with increasing turbulence (i.e., Reynolds number, $Re$), when considering the simulations that are included in the empirical fit in Figure~\ref{fig:scaling}. Ultimately, whether the properties of the dynamically-induced stable regions found in this study continue to apply at more realistic physical conditions must be tested by future numerical simulations.  

Our results suggest a number of implications that may improve understanding of the structure and dynamics at the top of Earth's outer core. A mixture of stable and unstable regions below the CMB allows radial motion at the top of the core, which seems a likely requirement to explain observed geomagnetic secular variation \citep[e.g.,][]{Amit2014CanVariation,lesur_2015,gastine_2020}, while also preserving thermochemical anomalies that might bear a seismically observable signature. Thermally-driven dynamo simulations with strong and thick RILs have generated magnetic fields that reproduced the large-scale morphology and secular variation of the recent geomagnetic field \citep{mound_2023}, and future work should test this result for the thermochemical case. Seismic studies have begun to investigate regional variations in compressional wave velocity \citep{ma2024seismic}, and our results provide testable predictions for the anomalous structures. Future studies could test these predictions by comparing SmKS body waves that sample distinct regions of the core, including beneath Africa, the Pacific, and the polar regions. Finally, the strength of the stable regions depends on the core's chemistry, providing a means to infer the range of candidate core compositions that are consistent with the predicted value of $\Nsq$.

\begin{acknowledgements}
SN, CJD and JEM are supported by Natural Environment Research Council research grant NE/W005247/1. ATC and CJD are supported by NE/V010867/1. This work used the ARCHER2 UK National Supercomputing Service (\href{http://www.archer2.ac.uk}{http://www.archer2.ac.uk}) and ARC4 (\href{https://arcdocs.leeds.ac.uk/welcome.html}{https://arcdocs.leeds.ac.uk/welcome.html}), part of the High-Performance Computing facilities at the University of Leeds, UK. We gratefully acknowledge two anonymous reviewers and the editor for their suggestions that improved this work.
\end{acknowledgements}

%%%%%%%%%%%%%%%%%%%%%%%%%%%%%%%%%%%%%%%%%%%%%%%%%%%%%%%%%%%%
\section*{Competing interests}\label{sec:COI}
%%%%%%%%%%%%%%%%%%%%%%%%%%%%%%%%%%%%%%%%%%%%%%%%%%%%%%%%%%%%
The authors declare that they have no competing financial interests or personal relationships that could have influenced the work reported in this paper.

%%%%%%%%%%%%%%%%%%%%%%%%%%%%%%%%%%%%%%%%%%%%%%%%%%%%%%%%%%%%
\section*{Data availability}
%%%%%%%%%%%%%%%%%%%%%%%%%%%%%%%%%%%%%%%%%%%%%%%%%%%%%%%%%%%%
The data that support the findings of this study are openly available in the National Geoscience Data Centre (NGDC) at \href{https://doi.org/10.5285/74c2ed9d-6ab4-4d24-863d-5991afbe84ce}{https://doi.org/10.5285/74c2ed9d-6ab4-4d24-863d-5991afbe84ce}

\appendix

\section{Supplementary Information}\label{app:SI}
\setcounter{figure}{0}
\renewcommand{\thefigure}{A\arabic{figure}}

\setcounter{table}{0}
\renewcommand{\thetable}{A\arabic{table}}

\setcounter{equation}{0}
\renewcommand{\theequation}{A\arabic{equation}}
\subsection{Detailed methodology}\label{sec:methods_sup}

%Adapted from Mound and Davies 2017, Monville et al. 2019, Trumper et al. 2012 etc.
%We employ a numerical model of convection of a Boussinesq fluid confined within a rotating spherical shell. 
%The fluid is a mixture of light components dissolved in a comparatively heavy liquid (e.g. Oxygen mixed in liquid iron in Earth's outer core). The relevant physical properties of the mixture are the kinematic viscosity, $\nu$, the thermal and chemical diffusivities, $\kappa_T$ and $\kappa_\xi$,  the coefficients of thermal and compositional expansion, $\alpha_T$ and $\alpha_\xi$. The thermal diffusivity is defined as $\kappa_T=k_T/\rho_o c_p$, where $k_T$ is the thermal conductivity, $\rho_o$ is the reference density, and $c_p$ is the specific heat capacity of the mixture. A spherical coordinate system $(r,\theta,\phi)$ is used to represent the domain bounded by the inner and outer boundaries, $r_i$ and $r_o$, respectively. The whole system rotates with a constant angular velocity $\boldsymbol{\Omega}=\Omega\boldsymbol{\hat{z}}$ about the vertical axis. 
The governing equations for the conservation of mass, momentum, energy, and chemical composition can be written as follows:

\begin{equation}\label{eqn:continuity_d}
   \boldsymbol{\nabla\cdot u}=0,
\end{equation}

\begin{equation}\label{eqn:momentum_d}
\begin{split} 
\left(\frac{\partial \boldsymbol{u}}{\partial t}+    (\boldsymbol{u}\cdot\boldsymbol{\nabla})\boldsymbol{u}\right)+2(\boldsymbol{\Omega}\times\boldsymbol{u})  =-\boldsymbol{\nabla}P+\frac{\trho}{\rho_o}\boldsymbol{g}+
\nu{\nabla^{2}}\boldsymbol{u},
\end{split}
\end{equation}

\begin{equation}\label{eqn:energy_d}
\frac{\partial T}{\partial t}+    (\boldsymbol{u}\cdot\boldsymbol{\nabla})T =
\kappa_{T}{\nabla^{2}}T+S_T,
\end{equation}

\begin{equation}\label{eqn:composition_d}
\frac{\partial \xi}{\partial t}+    (\boldsymbol{u}\cdot\boldsymbol{\nabla})\xi =
\kappa_{\xi}{\nabla^{2}}\xi+S_{\xi},
\end{equation}

The modified pressure, $P$, includes the centrifugal potential. Gravity varies linearly with the radius such that $\boldsymbol{g}= -(g_o/r_o)\boldsymbol{r}$, where $g_o$ is the reference gravitational acceleration at the CMB ($r=r_o$). We use the Boussinesq approximation where the variations of the density $\trho$ due to the temperature $\tT$ and concentration of light elements $\tC$ are only taken in the buoyancy force. Following a linear equation of state, we get 

\begin{align}\label{eqn:bousensq}
    \frac{\trho}{\rho_o}=\frac{\rho-\rho_o}{\rho_o}=-\alpha_T(\tT-T_o)-\alpha_\xi(\tC-\xi_o),
\end{align}
by assuming $|\rho-\rho_o|/\rho_o \ll 1$, where $T_o,\xi_o,\rho_o$ are the reference values at $r=r_o$. Because the scalar magnitudes have no dynamic significance (unlike the gradients of these scalar quantities), we set $(T_o, \xi_o) = (0, 0)$. In this equation of state \ref{eqn:bousensq}, $\tT$ is the departure from the destabilising background superadiabatic temperature profile ($T_c$, henceforth referred to as the conductive state), whereas $\tC$ is the departure from the compositional reference barodiffusive profile $\xi_c$ \citep{davies_2011}, as expressed below: 

\begin{equation}\label{eqn:tempdef}
    \begin{split}
    \tT=T-T_c, \\
    \tC=\xi-\xi_c.
    \end{split}
\end{equation}

Fixed-flux thermal boundary conditions are imposed at the inner and outer boundaries such that the total radial heat flow is equal through the inner and outer surfaces ($\boldsymbol{Q}_{T,i}=\boldsymbol{Q}_{T,o}$). The temperature gradient at the boundaries is expressed as $\boldsymbol{\nabla}T_{c}=-(\beta_{T}/r^{2})\boldsymbol{\hat{r}}$ and related to the heat flux at the inner core boundary (ICB) through Fourier Law as, $\boldsymbol{Q}_{T,i}=4\pi r_{i}^{2}\boldsymbol{q_{T}^{avg}}=4\pi r_{i}^{2}(-k_{T}\boldsymbol{\nabla}T_{c})=4\pi k_{T} \beta_T \boldsymbol{\hat{r}}$. The thermal convection is maintained entirely by this heat flow, and therefore, we set $S_T=0$ in equation  \ref{eqn:energy_d}. Without a heat source, the thermal conduction equation can be expressed as

\begin{equation}\label{eqn:conduction}
\frac{\kappa_T}{r^2}\frac{d}{dr} \left( r^2\frac{dT_c}{dr} \right) = 0,
\end{equation}
subjected to the boundary conditions
\begin{align}\label{eqn:conduction_bc}
\left( \frac{dT_c}{dr} \right)_{r=r_i} = -\frac{\beta_T}{r_i^{2}},\qquad
\left( \frac{dT_c}{dr} \right)_{r=r_o} = -\frac{\beta_T}{r_o^{2}}.
\end{align}

The solution is
\begin{align}\label{eqn:conduction_sol}
\frac{dT_c}{dr} = -\frac{\beta_T}{r^{2}},\qquad
T_c(r) = \frac{\beta_T}{r}+c_1.
\end{align}

Here, the constant of integration $c_1$ is not constrained by the boundary conditions. It can be set to zero without any loss of generality, as only the gradient has dynamic significance. The temperature drop across the shell for purely thermal conduction is

\begin{align}\label{eqn:tempdrop}
\Delta T_c= \beta_T\left(\frac{1}{r_i}-\frac{1}{r_o}\right)=\frac{\beta_Th}{r_ir_o},
\end{align}
where $h=r_o-r_i$ is the shell gap.  

We assume zero compositional flux at the CMB (i.e. $\boldsymbol{{Q}}_{\xi,o}=\boldsymbol{0}$), while imposing fixed flux conditions at the inner boundary. To ensure stationary solutions, we assume that the flux from the inner core is balanced by a spatially homogeneous sink ( $S_\xi$ ) that maintains the global balance of lighter elements. With a sink term, the light element diffusion equation can be expressed as

\begin{equation}\label{eqn:massdiffusion}
\frac{\kappa_\xi}{r^2}\frac{d}{dr} \left( r^2\frac{d\xi_c}{dr} \right) +S_\xi=0,
\end{equation}

subjected to the boundary conditions

\begin{align}\label{eqn:massdiff_bc}
\left( \frac{d\xi_c}{dr} \right)_{r=r_i} = -\frac{\beta_\xi}{r_i^{2}},\qquad
\left( \frac{d\xi_c}{dr} \right)_{r=r_o} = 0.
\end{align}

The solution is

\begin{align}\label{eqn:massdiff_sol}
\frac{d\xi_c}{dr} = \frac{\beta_\xi}{r_o^3-r_i^3}(r-r_o^3/r^2),\qquad
\xi_c(r) = \frac{\beta_\xi}{2(r_o^3-r_i^3)}(r^2+2r_o^{3}/r)+c_2,
\end{align}

where, $S_\xi=-3\beta_\xi\kappa_\xi/(r_o^{3}-r_i^3)$ and the constant of integration $c_2$ is not constrained by the boundary conditions. 

%We get,
%
%\begin{equation}\label{eqn:diff_profile}
%\frac{d\xi_c}{dr} = \frac{\beta_\xi}{r_o^3-r_i^3}(r-r_o^3/r^2),\qquad
%\xi_c(r) = \frac{\beta_\xi}{2(r_o^3-r_i^3)}(r^2+2r_o^{3}/r)+c_2
%\end{equation}

\subsubsection{Non-dimensional governing equations}\label{sec:gov_eq_nd}

We proceed by non-dimensionalising equations \ref{eqn:continuity_d}-\ref{eqn:composition_d}  using the shell gap $h$ as length scale, the viscous diffusion time, $h^2/\nu$ , as time-scale, $\beta_T/h$ as temperature scale and $\beta_\xi/h$ as the scale of the compositional field. For these choices, the non-dimensional equations become
\begin{equation}\label{eqn:continuity_nd}
   \boldsymbol{\nabla^{*}\cdot u^{*}}=0,
\end{equation}

\begin{equation}\label{eqn:momentum_nd}
%\begin{split} 
\frac{\partial \boldsymbol{u^{*}}}{\partial t^{*}}+    (\boldsymbol{u^{*}}\cdot\boldsymbol{\nabla^{*}})\boldsymbol{u^{*}}+\frac{1}{E}\left(\boldsymbol{\hat{z}}\times\boldsymbol{u^{*}}\right)  =-\boldsymbol{\nabla^{*}}P^{*}+%\\
\left(\frac{Ra_{T}}{Pr_T}\tT^{*}+\frac{Ra_{\xi}}{Pr_\xi}\tC^{*}\right)\boldsymbol{r}^{*}+
{\nabla^{*2}}\boldsymbol{u^{*}},
%%\end{split}
\end{equation}
 
\begin{equation}\label{eqn:energy_nd}
\frac{\partial T^{*}}{\partial t^{*}}+    (\boldsymbol{u}^{*}\cdot\boldsymbol{\nabla}^{*})T^{*} = \frac{1}{Pr_{T}}{\nabla^{*2}}T^{*},
\end{equation}

\text{and}
 
\begin{equation}\label{eqn:composition_nd}
\frac{\partial \xi^{*}}{\partial t^{*}}+    (\boldsymbol{u}^{*}\cdot\boldsymbol{\nabla}^{*})\xi^{*} =
\frac{1}{Pr_{\xi}}{\nabla^{*2}}\xi^{*}-\frac{1}{Pr_{\xi}}\frac{3}{r^{*3}_o-r^{*3}_i}.
\end{equation}

\subsubsection{Non-dimensional background states}\label{sec:background_nd}

Non-dimensionalising the background profiles in equations \ref{eqn:conduction_sol} and \ref{eqn:massdiff_sol}, we get

\begin{align}\label{eqn:conduction_sol_nd}
\frac{dT^{*}_c}{dr^{*}} = -\frac{1}{r^{*2}},\qquad
T^{*}_c(r^{*}) = \frac{1}{r^{*}}+c_1,\qquad
\end{align}

\begin{align}\label{eqn:conduction_temp_diff}
\Delta T^{*}_c = T^{*}_c(r_i^{*})-T^{*}_c(r_o^{*})=\frac{1}{r_i^{*}}-\frac{1}{r_o^{*}}=\frac{(1-\eta)^{2}}{\eta},
\end{align}
where $\eta=r_i/r_o$ is the radius ratio fixed at $\eta=0.35$ in our study. For the compositional background profile, we obtain
\begin{align}\label{eqn:diff_profile_nd}
%\begin{split}
\frac{d\xi^{*}_c}{dr^{*}} = \frac{1}{r_o^{*3}-r_i^{*3}}(r^{*}-r_o^{*3}/r^{*2}),\\
\xi^{*}_c(r^{*}) = \frac{1}{2(r_o^{*3}-r_i^{*3})}(r^{*2}+2r_o^{*3}/r^{*})+c_2 ,   
%\end{split}
\end{align}

\begin{align}\label{eqn:diffusion_comp_diff}
\begin{split}
\Delta \xi^{*}_c= \xi^{*}_c(r_i^{*})-\xi^{*}_c(r_o^{*})=\frac{(1-\eta)^{2}}{\eta^2}\frac{\eta(\eta+2)}{2(1+\eta+\eta^{2})}.    
\end{split}
\end{align}

\subsubsection{Diagnostic quantities}\label{sec:diagnostics}
A few diagnostic quantities are tabulated in appendix \ref{app:diagnostics} reflecting the time-averaged global behaviour of the simulations, as well as properties of the stratified regions. The global flow velocity, convective heat and light element flux can be represented by the Reynolds number ($Re=Uh/\nu=U^{*}=\sqrt{2KE/V_s}$), thermal and chemical Nusselt numbers ($\NuT=\Delta T^{*}/ \Delta T_{c}^{*}$ and $\NuC=\Delta \xi^{*}/ \Delta \xi^{*}_c$), respectively. Here, $U^{*}$ is the non-dimensional characteristic velocity, $V_s$ is the shell volume, $KE=1/2\iiint_{V_s}\boldsymbol{u}^{*}.\boldsymbol{u}^{*} dV$ is the kinetic energy integral, $\Delta T^{*}$ and $\Delta \xi^{*}$ are the difference in surface and time average temperature and light element concentration between inner and outer radius, with the drop in conductive temperature and concentrations (i.e. $\Delta T_{c}^{*}$ and $\Delta \xi^{*}_c$) given in equations \ref{eqn:conduction_temp_diff} and \ref{eqn:diffusion_comp_diff}, respectively.

\begin{figure}[h!]%% placement specifier
\centering
\includegraphics[width=\linewidth,trim={6cm 3cm 6cm 2cm},clip]{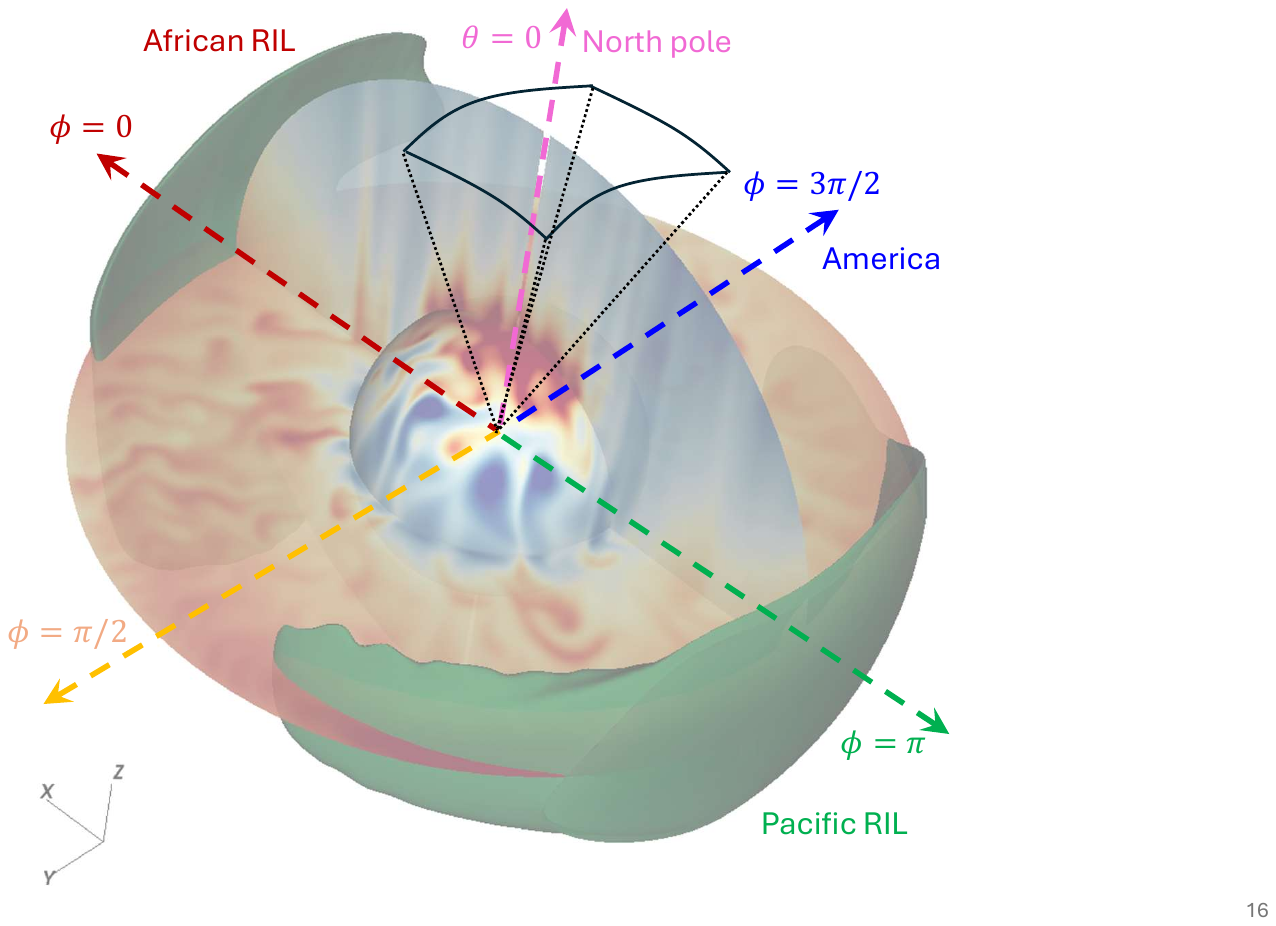}
\caption{Geometric orientation of the RILs with respect to the Earth's surface.}\label{fig:geometry}
\end{figure}

We represent the radial gradient of density in terms of the Brunt-V\"{a}is\"{a}l\"{a} (BV) frequency defined as

 \begin{equation}\label{eqn:Nsq_d}    
    N^2=-\frac{g}{\rho_o}\frac{\partial\rho}{\partial r}.
 \end{equation}

Combining equations \ref{eqn:Nsq_d} and \ref{eqn:bousensq}, with $(T_o, \xi_o) = (0, 0)$ and $g/g_o=r/r_o$ gives

\begin{equation}\label{eqn:Nsq_d2}   
    N^2=g_o\left(\frac{r}{r_o}\right)\left(\alpha_T\Trd+\alpha_\xi\Crd\right);
\end{equation}

 The non-dimensional form of $N^2$ is given by

\begin{equation}\label{eqn:Nsq_nd_SI}  
    \Nsq = r^{*}E^2\left(\GrT\Trnd+\GrC\Crnd\right)\equiv \NTsq+\NCsq,
\end{equation}

where $\NTsq$ and $\NCsq$ are the BV frequencies estimated from the radial gradients of thermal and chemical fields. Henceforth, we drop the asterisk from the superscripts of the non-dimensional quantities used in the following section.

\subsubsection{BV frequency profiles of thermal and chemical fields}\label{sec:NTsq}
\begin{figure*}[h!]%% placement specifier
\centering
Chemically dominated$\color{Blue} \boldsymbol{(\square)}$\hspace{5cm} Thermally dominated$\color{BrickRed} \boldsymbol{(\square)}$\\
\vspace{5mm}
\begin{overpic}[width=0.48\linewidth,trim={0cm 0cm 0cm 0cm},clip]{\figfile{NTsq_depth_Pm=0_Pr_T=1_Pr_C=10_q=0.0_E=1e-5_Ra_T=90_Ra_C=30000}}
\put(0,80){$(a)$}
\end{overpic}
\begin{overpic}[width=0.48\linewidth,trim={0cm 0cm 0cm 0cm},clip]{\figfile{NTsq_depth_Pm=0_Pr_T=1_Pr_C=10_q=0.0_E=1e-5_Ra_T=1200_Ra_C=1000}}
\put(0,80){$(b)$}
\end{overpic}\\
\begin{overpic}[width=0.48\linewidth,trim={0cm 0cm 0cm 0cm},clip]{\figfile{NCsq_depth_Pm=0_Pr_T=1_Pr_C=10_q=0.0_E=1e-5_Ra_T=90_Ra_C=30000}}
\put(0,80){$(c)$}
\end{overpic}
\begin{overpic}[width=0.48\linewidth,trim={0cm 0cm 0cm 0cm},clip]{\figfile{NCsq_depth_Pm=0_Pr_T=1_Pr_C=10_q=0.0_E=1e-5_Ra_T=1200_Ra_C=1000}}
\put(0,80){$(d)$}
\end{overpic}
\caption{Variation of $\NTsq$ (a,b) and $\NCsq$ (c,d) with depth below the CMB, near the top of the core, averaged around various locations for models for $\qstar=0$ with compositionally dominated convection at $\tRaT=90$ and $\tRaC=30000$ (a,c), and thermally dominated convection at $\tRaT=1200$ and $\tRaC=300$ (b,d).}\label{fig:homog_prof}
\end{figure*}

\begin{figure*}[h!]%% placement specifier
\centering
Chemically dominated$\color{Blue} \boldsymbol{(\square)}$\hspace{5cm} Thermally dominated$\color{BrickRed} \boldsymbol{(\square)}$\\
\vspace{5mm}
\begin{overpic}[width=0.48\linewidth,trim={0cm 0cm 0cm 0cm},clip]{\figfile{NTsq_depth_Pm=0_Pr_T=1_Pr_C=10_q=5.0_E=1e-5_Ra_T=90_Ra_C=30000}}
\put(0,80){$(a)$}
\end{overpic}
\begin{overpic}[width=0.48\linewidth,trim={0cm 0cm 0cm 0cm},clip]{\figfile{NTsq_depth_Pm=0_Pr_T=1_Pr_C=10_q=5.0_E=1e-5_Ra_T=1200_Ra_C=1000}}
\put(0,80){$(b)$}
\end{overpic}
\begin{overpic}[width=0.48\linewidth,trim={0cm 0cm 0cm 0cm},clip]{\figfile{NCsq_depth_Pm=0_Pr_T=1_Pr_C=10_q=5.0_E=1e-5_Ra_T=90_Ra_C=30000}}
\put(0,80){$(c)$}
\end{overpic}
\begin{overpic}[width=0.48\linewidth,trim={0cm 0cm 0cm 0cm},clip]{\figfile{NCsq_depth_Pm=0_Pr_T=1_Pr_C=10_q=5.0_E=1e-5_Ra_T=1200_Ra_C=1000}}
\put(0,80){$(d)$}
\end{overpic}
\caption{Variation of $\NTsq$ (a,b) and $\NCsq$ (c,d) with depth below the CMB, near the top of the core, in the heterogeneous models ($\qstar=5$), averaged around various locations for compositionally dominated convection at $\tRaT=90$ and $\tRaC=30000$ (a,c), and thermally dominated convection at $\tRaT=1200$ and $\tRaC=300$ (b,d).}\label{fig:tomog_prof}
\end{figure*}
In Figures~\ref{fig:homog_prof} and Figures~\ref{fig:tomog_prof}, the locally averaged radial profiles for $\NTsq$ (top rows) and $\NCsq$ (bottom rows) are plotted to investigate the properties of the thermally and/or chemically stable regions, while their sum $\Nsq$ (equation \ref{eqn:Nsq_nd_SI}) is provided in the main text (Figure \ref{fig:Nsq_prof}). In these figures, the profiles denoted as ``Africa'' and ``Pacific'' are averaged from the radial lines whose outer edge falls within the area over the CMB with an imposed stable temperature gradient (i.e. $\Trinline(r=r_o)>0$), centered around the positive ($\phi=0$) and negative $x$-axis ($\phi=\pi$), respectively, for the heterogeneous models (see Figure \ref{fig:geometry}). The same area beneath Africa and the Pacific has been used for the homogeneous models to average the radial profiles consistently, although there is no imposed stable temperature gradient at the CMB. Two more spatially averaged profiles are calculated at the equatorial regions, one averaged around $\phi=3\pi/2$ (America) and another around $\theta=0$ (North Pole), from radial profiles with their outer edge falling within an arbitrary spherical surface on the CMB, that creates a solid angle of $\pi/2$ with the centre (Figure~\ref{fig:geometry}). In general, the dashed arrows in Figure~\ref{fig:geometry}, colour-coded for correspondence with figures \ref{fig:homog_prof} and \ref{fig:tomog_prof} in this appendix and Figure~\ref{fig:Nsq_prof} in the main text, indicate the location around which the radial profiles are extracted for spatial averaging. We note here that all these locally averaged profiles in these plots, as well as the globally averaged profile (black line), are extracted from time-averaged fields of the scalar gradients. The profiles around the Indian Ocean (orange line) and the South Pole are similar to American and North Polar profiles, respectively, and hence are omitted from the figures for clarity.

\subsubsection{Radial velocity field maps}\label{sec:ur_radplane}
\begin{figure*}[h!]%% placement specifier
\centering
Chemically dominated $\color{Blue} \boldsymbol{(\square)}$\hspace{5cm} Thermally dominated $\color{BrickRed} \boldsymbol{(\square)}$\\
\vspace{10mm}
\begin{overpic}[width=0.47\linewidth,trim={20cm 13cm 18cm 15cm},clip]{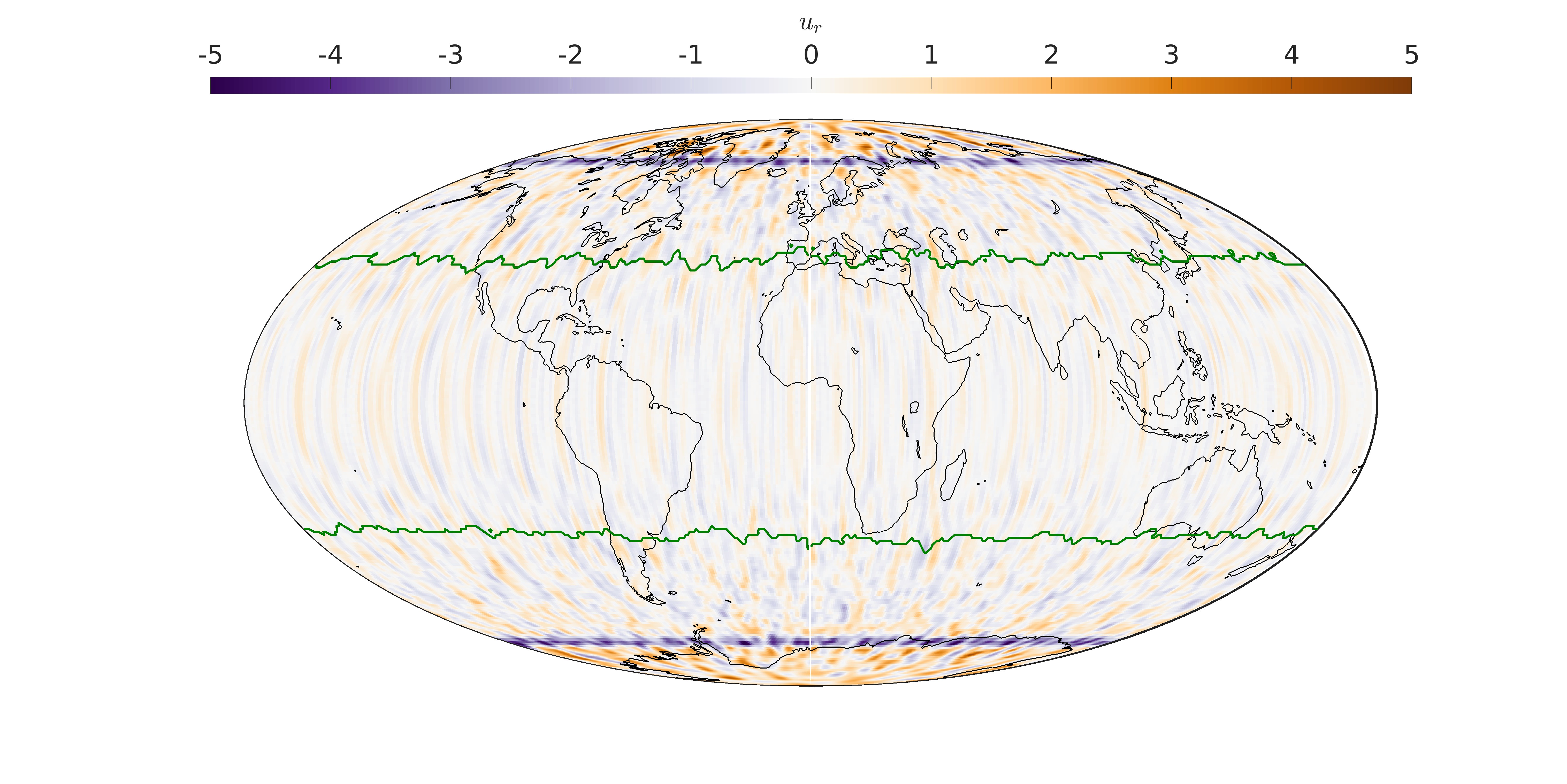}
\put(5,56){$(a)\ \textrm{Homogeneous}$}
\end{overpic}
\begin{overpic}[width=0.48\linewidth,trim={20cm 13cm 18cm 4cm},clip]{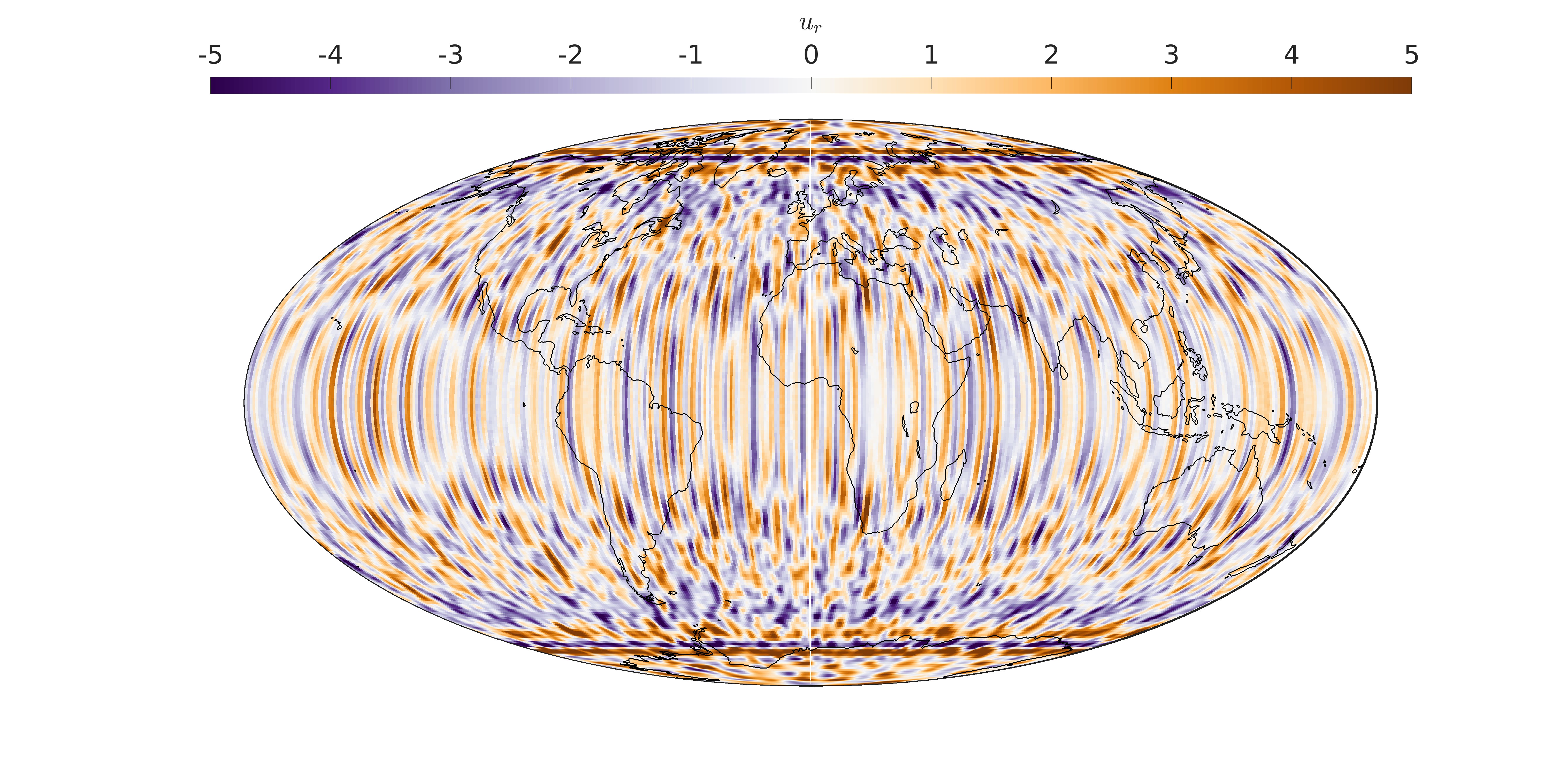}
\put(0,53){$(b)\qquad\qquad\qquad\qquad\quad u_{r}$}
\end{overpic}
\vspace{1mm}
\vspace{4mm}
\begin{overpic}[width=0.47\linewidth,trim={20cm 13cm 18cm 15cm},clip]{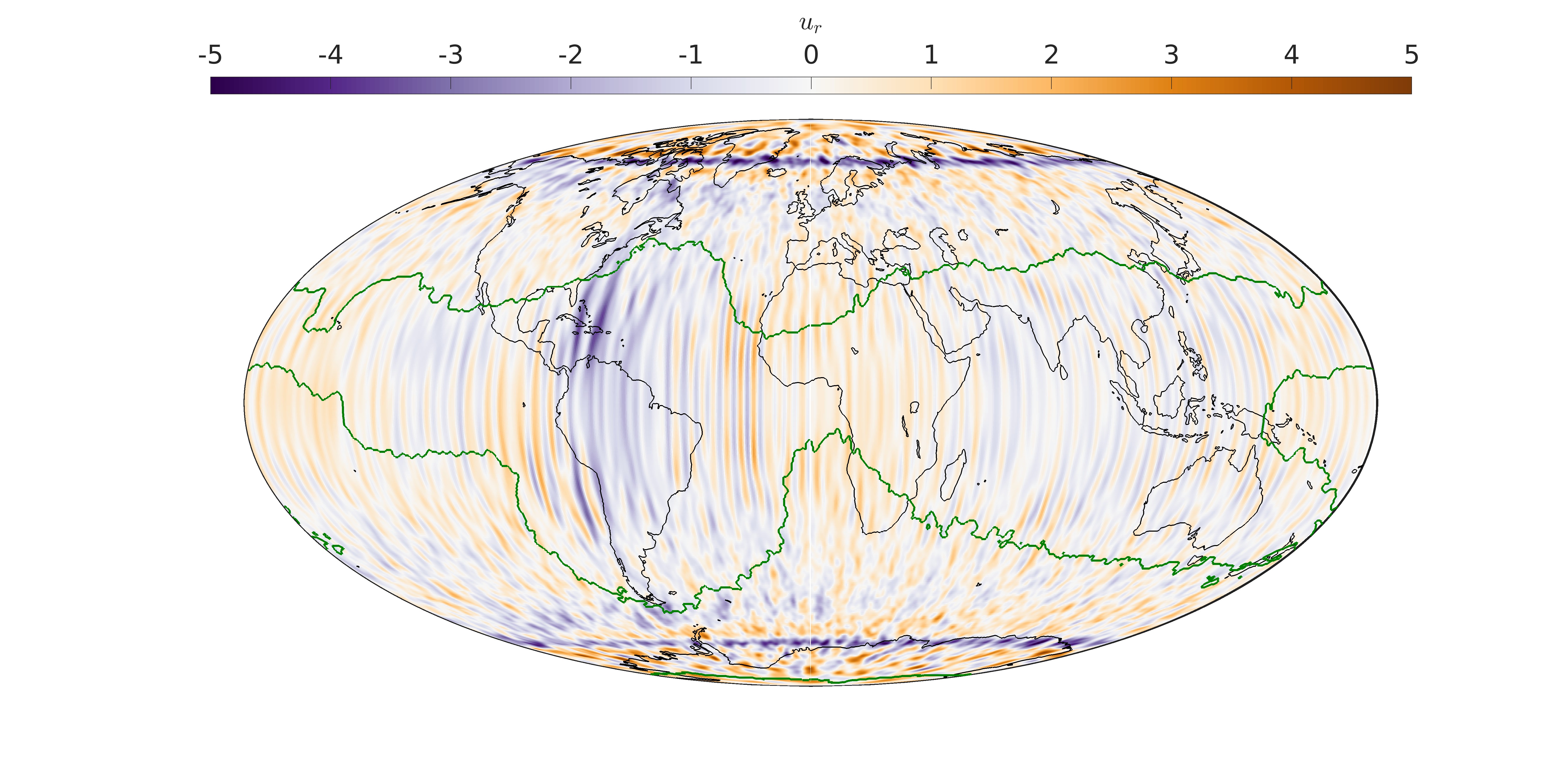}
\put(0,44){$(c)\ \textrm{Heterogeneous}$}
\end{overpic}
\begin{overpic}[width=0.47\linewidth,trim={20cm 13cm 18cm 15cm},clip]{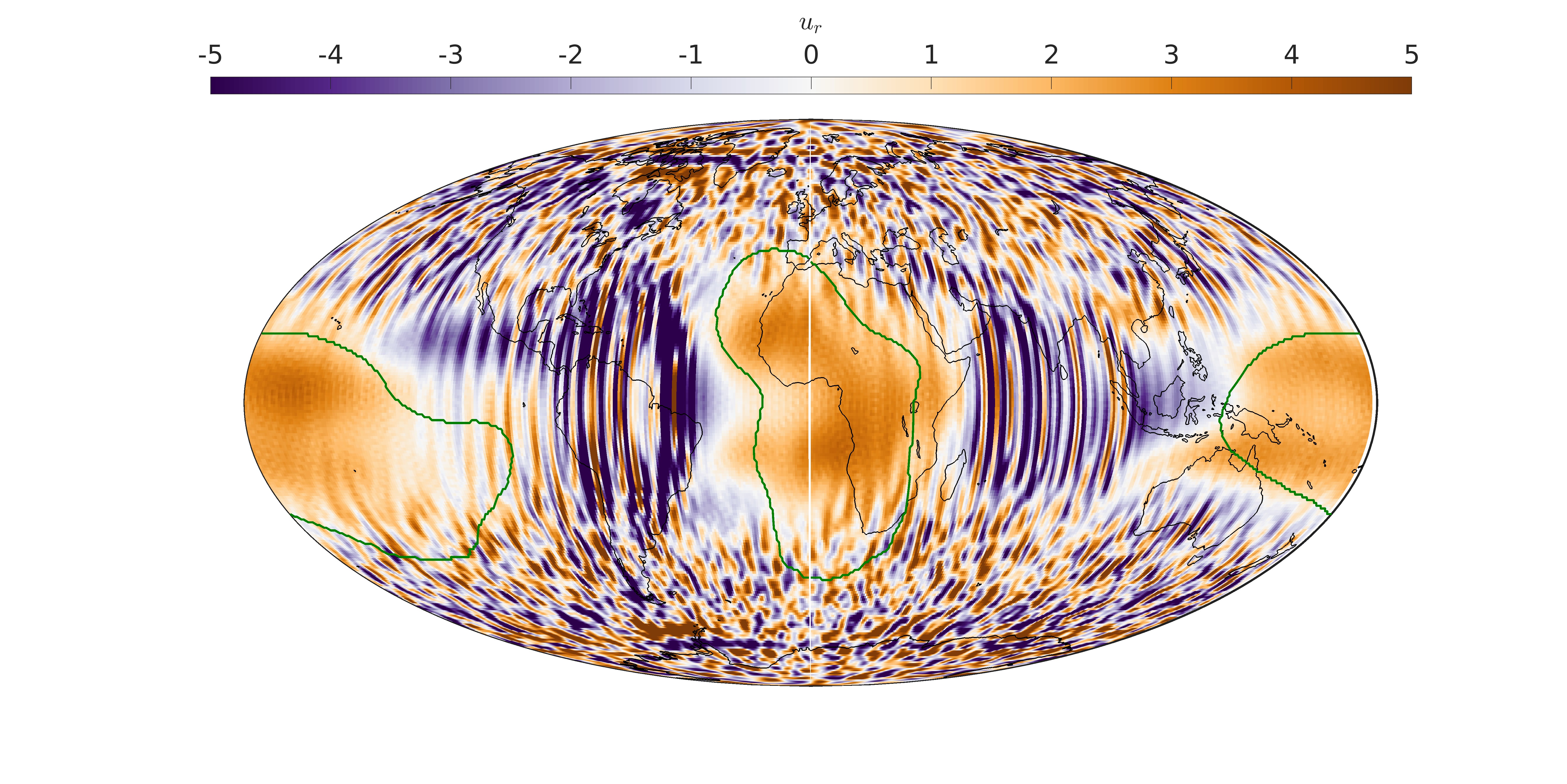}
\put(0,44){$(d)$}
\end{overpic}
\caption{Time-averaged radial velocity map $\sim100$ km below the CMB for (a,b) homogeneous and (c,d) heterogeneous cases. For each set, the Rayleigh numbers are (a,c) $\tRaT=90$ and $\tRaC=30000$ and (b,d) $\tRaT=1200$ and $\tRaC=300$ ( same cases as Figures \ref{fig:homog} and \ref{fig:tomog} in the main text). The green lines show the boundary of stratified regions with $\Nsq=0$.}\label{fig:vel_radmap}
\end{figure*}

Maps of the time-averaged radial velocity near the top of the core (Figure \ref{fig:vel_radmap}) complement the visualisation of instantaneous radial flow (Figures \ref{fig:homog} and \ref{fig:tomog} in the main text). The zero CMB compositional flux condition at the CMB suppresses motion near the top of the core in the chemically dominated cases, compared to the thermally dominated cases. The RILs in the thermally dominated heterogeneous case (Figure \ref{fig:vel_radmap}(d)) suppress small-scale motion and are characterised by a broad upwelling flow similar to the findings of \citet{mound_2019}. 

\subsubsection{Global balance of light elements}\label{sec:comp_bal}

\begin{figure*}[h!]%% placement specifier
\vspace{4mm}
\centering
\begin{overpic}[width=0.6\linewidth,trim={0cm 1cm 0cm 0cm},clip]{\figfile{glob_bal_dsource_Pm=0_Pr_T=1_Pr_C=10_q=0.0_E=1e-5_Ra_T=0_Ra_C=100000}}
\put(-2,52){$(a)$}
\end{overpic}

\vspace{2mm}

\begin{overpic}[width=0.6\linewidth,trim={0cm 0cm 0cm 1.5cm},clip]{\figfile{glob_bal_dsource_cmb_Pm=0_Pr_T=1_Pr_C=10_q=0.0_E=1e-5_Ra_T=0_Ra_C=100000}}
\put(-2,50){$(b)$}
\end{overpic}
\caption{Radial dependence of the terms in equation \ref{eqn:comp_bal2} for varying $\tRaC$ for four purely compositional convection simulations at $\tRaC\in\{100,\ 1000,\ 10000,\ 100000 \}$. The full radial extent is shown in (a) and zoomed in near the boundaries in (b). The convective (coloured solid lines) and diffusive (coloured dashed lines) flux terms are coloured by $\tRaC$ (colourbar). The solid teal line in (a) indicates the sink term, which is independent of $\tRaC$. The black horizontal line represents the total sum, while the difference between the total and the sink term is denoted as $\Delta\textrm{sink}$ (black dash-dot line)}.\label{fig:comp_bal}
\end{figure*}

We study the mechanism of formation of chemically stable regions at the top of the core from the balance of light elements in equation \ref{eqn:composition_nd}. Integrating the equation over a spherical shell of inner radius $r_i$ and a varying outer radius $r$, with shell volume, $V_{s}=(4/3)\pi (r^{3}-r_{i}^{3})$, we get

%\begin{equation}\label{eqn:comp_int}
\begin{align}\label{eqn:comp_bal}
\iiint_{V_{s}}\frac{\partial \xi}{\partial t}\,dV+\iiint_{V_{s}}(\boldsymbol{u}\cdot\boldsymbol{\nabla})\xi\,dV- \notag
\iiint_{V_{s}}\frac{1}{Pr_{\xi}}{\nabla^{2}\xi}\,dV \\
-\iiint_{V_{s}}\frac{S_{\xi}}{Pr_{\xi}}\,dV=0,
\end{align}
%\end{equation}
where $S_{\xi}=-3/(r^{3}_o-r^{3}_i$) is the non-dimensional volumetric sink. 

Using the divergence theorem the second term reduces to \(\oint_{S} \langle u_{r}\xi\rangle \,dS=4\pi r^{2}\langle u_{r}\xi\rangle\), while the third term becomes $-4\pi r^{2}(\partial\xi/\partial r)-4\pi$. After some manipulation and rearrangement, and using non-dimensional sink $S_\xi=-3/(r^{3}_{o}-r^3_{i})$ we get

\begin{gather}\label{eqn:comp_bal2}
\underbrace{\PrC\iiint_{V_{s}}\frac{\partial \xi}{\partial t}\,dV}_{\textrm{time derivative}}
\underbrace{+4\pi r^{2}\PrC\langle u_{r}\xi\rangle}_{\textrm{convective flux}}
\underbrace{-4\pi r^{2}\Crd}_{\textrm{diffusive flux}}
\underbrace{+4\pi\frac{r^{3}-r^3_{i}}{r^{3}_{o}-r^3_{i}}}_{\textrm{sink}}=\underbrace{4\pi}_{\textrm{total}}.
\end{gather}
%\end{equation}
We note here that the integral of the time derivative also goes to zero while considering the full shell volume (i.e. $r=r_{o}$), demonstrating global conservation of light elements: 

\begin{gather}\label{eqn:comp_bal3}
\PrC\iiint_{V_{s}(r_i)}^{V_{s}(r_o)}\frac{\partial \xi}{\partial t}\,dV=0.
\end{gather}

The time-derivative term also drops out after time-averaging the terms for statistically stationary solutions, and hence is not shown in Figure~\ref{fig:comp_bal}. The radial profiles of the terms in equation \ref{eqn:comp_bal2} are demonstrated in Figure \ref{fig:comp_bal} and discussed in the main text (section \ref{sec:stable_region}). Notably, the relative contributions of the convective and diffusive flux terms in equation \ref{eqn:comp_bal2} (i.e., the second and third terms) vary with $\tRaC$, whereas the other terms are independent of that parameter.

%\clearpage

\subsubsection{Scaling of chemically-driven stratification}\label{sec:LEA_scaling}

\begin{figure*}[h!]%% placement specifier
\centering
\hspace{1mm}
\begin{overpic}[width=0.45\linewidth,trim={0cm 0cm 0cm 0cm},clip]{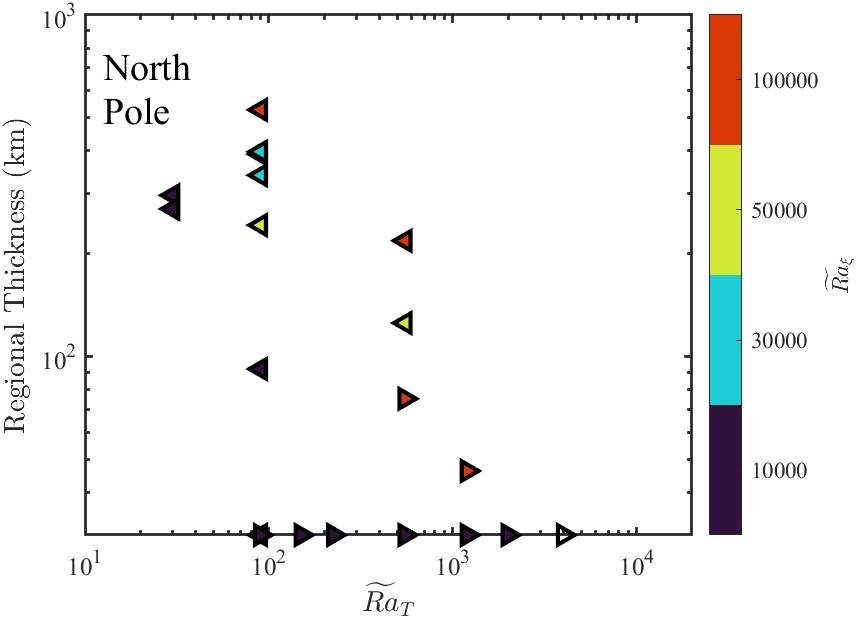}
\put(0,75){$(a)$}
\end{overpic}
%(b)
%\hspace{2mm}
\begin{overpic}[width=0.45\linewidth,trim={0cm 0cm 0cm 0cm},clip]{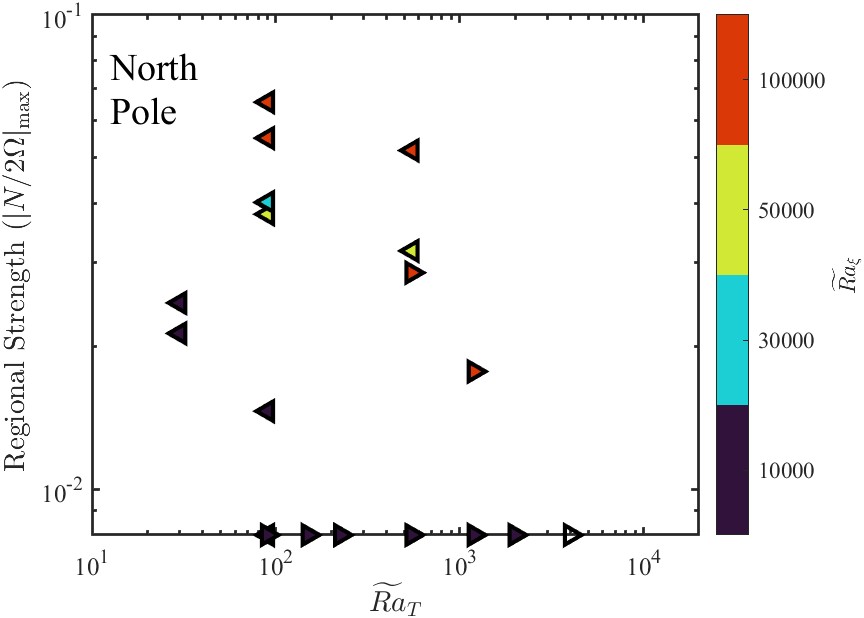}
\put(-2,75){$(b)$}
\end{overpic}\\
%
%(c)
\vspace{2mm}
\begin{overpic}[width=0.45\linewidth,trim={0cm 0cm 0cm 0cm},clip]{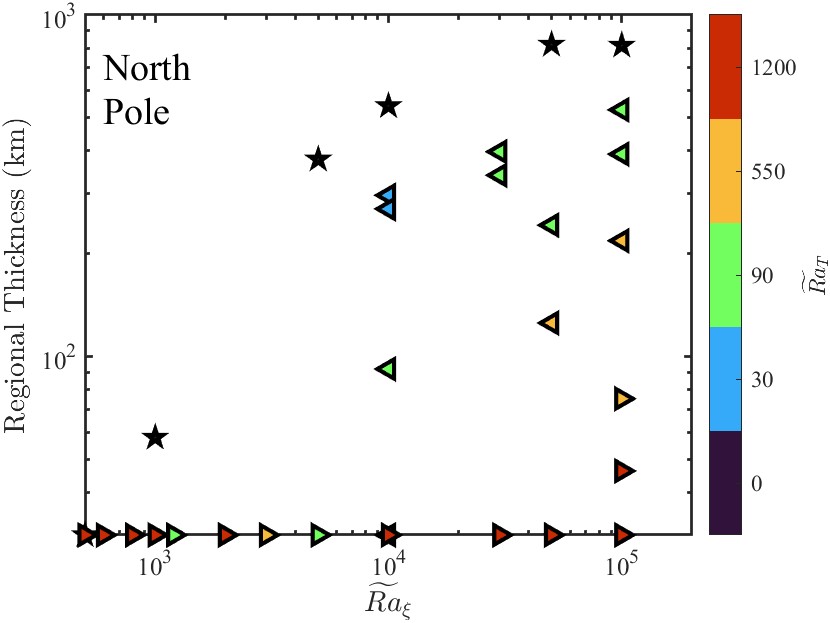}
\put(0,75){$(c)$}
\end{overpic}
%
%(d)
%\hspace{2mm}
\begin{overpic}[width=0.45\linewidth,trim={0cm 0cm 0cm 0cm},clip]{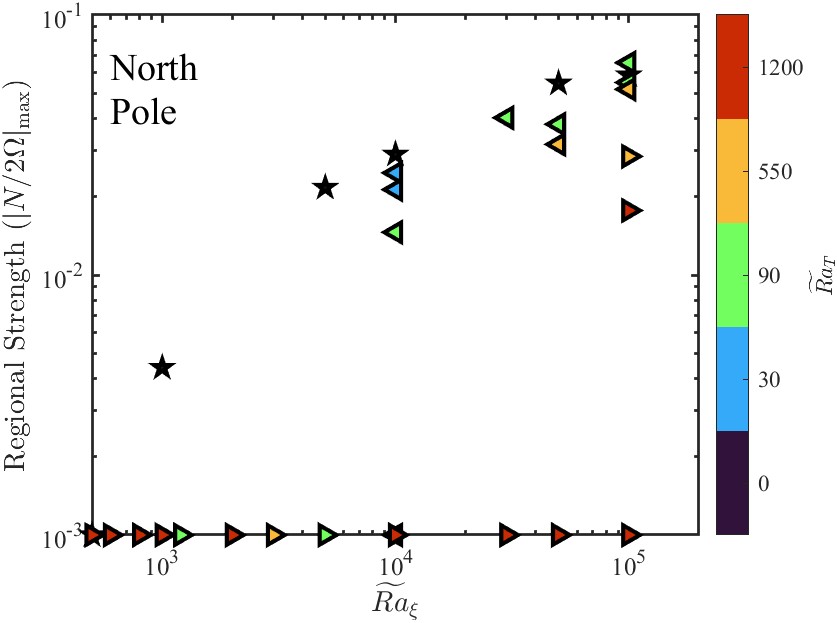}
\put(-2,75){$(d)$}
\end{overpic}
\caption{Thickness (a,c) and strength (b,d) of the stable regions averaged around the north pole plotted against thermal (a,b) and compositional (c,d) Rayleigh numbers. The symbol shapes indicate purely compositional (pentagrams) and thermochemical (left/right triangles) simulations. The thermochemical models are classified by the ratio of thermal to
chemical buoyancy $\FACT < 1$ (left triangles) and $\FACT > 1$ (right triangles), respectively. The cases with $\Nsq<0$ beneath the north pole are shown on the x-axis.}\label{fig:LEA_scaling}
\end{figure*}
The thickness and strength of the chemically-driven stratified regions beneath the north pole, as shown in Figure \ref{fig:LEA_scaling}, seem to increase with increasing $\tRaC$, and decrease with increasing $\tRaT$. 
%\clearpage

\subsubsection{Compositional fluxes in meridional plane}\label{sec:ur_radplane}
\begin{figure*}[h!]%% placement specifier
\vspace{1cm}
\centering
\begin{overpic}[width=\linewidth,trim={23cm 0cm 23cm 0cm},clip]{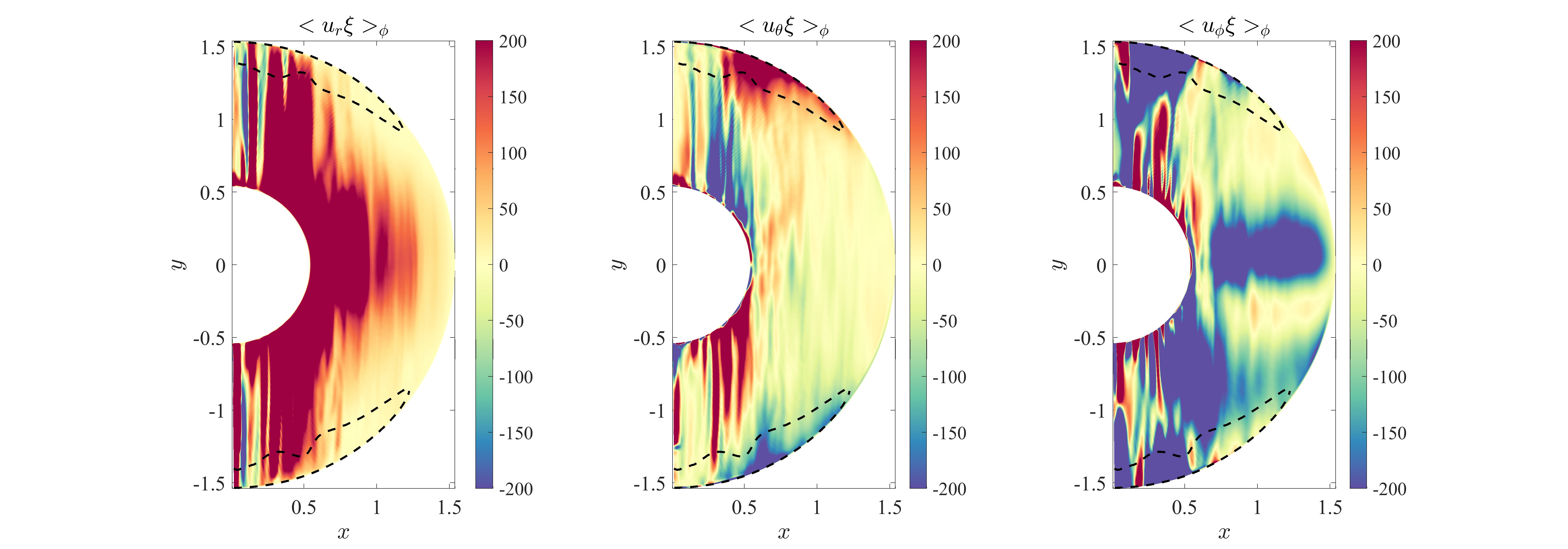}
\put(0,44){$(a)$\hspace{6cm}$(b)\qquad$\hspace{5.5cm}$(c)$}
\end{overpic}
\caption{Snapshot of longitudinally-averaged (a) radial (b) latitudinal and (c) longitudinal advective flux of composition for a homogeneous case at $\tRaT=90$ and $\tRaC=30000$ (same snapshot as Figure 2b in the main text). Here $\langle.\rangle_{\phi}$ indicates averaging in the longitudinal ($\phi$) direction.  The dashed lines show the time- and $\phi$-averaged boundary of the chemically stratified region.}\label{fig:comp_flux}
\end{figure*}
%\clearpage

The transport of light elements by the instantaneous flow is visualised from the longitudinally-averaged advective flux components for a chemically-dominated homogeneous simulation in Figure \ref{fig:comp_flux}.
%{They highlight the role of vertical transport of light elements near the tangent cylinder and the meridional circulation in the accumulation of light elements near the poles, followed by spreading towards lower latitudes.}

\subsubsection{Scaling of zonal flow}\label{sec:zonal}
\begin{figure*}[h!]%% placement specifier
\centering
\begin{overpic}[width=0.7\linewidth,trim={0cm 0cm 0cm 0cm},clip]{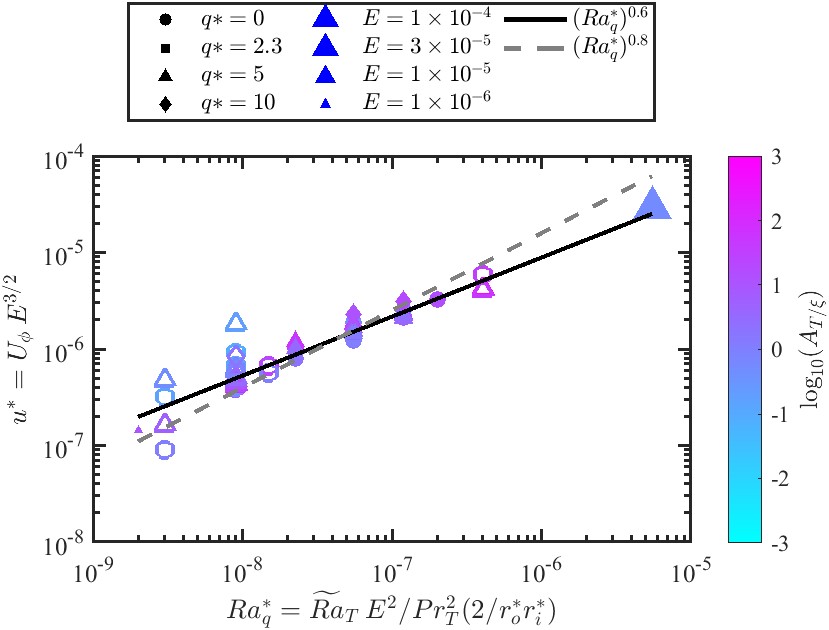}
\end{overpic}
\caption{Scaling of the zonal flow speed in thermochemical cases. The zonal velocity is calculated from the azimuthal kinetic energy $U_{\phi}=\sqrt{2(KE)_{\phi}/V_s}$ as defined in section \ref{sec:diagnostics}. Simulations with  $\FIC > 0.1$ (open symbols) are excluded from the fit.}\label{fig:zonal}
\end{figure*}

The scaling of zonal flow velocity magnitudes ($u^{*}$, which is rescaled to rotational units) with input convective power (expressed in terms of a rescaled Rayleigh number, $Ra^{*}_q=\tRaT E^2/\PrT^2(2/(r^*_o r^*_i)$)) in our thermochemical simulations is compared with an established scaling law for pure thermal convection proposed by \citet{aubert_2005} (Figure \ref{fig:zonal}). The theoretical scaling law, $u^{*}\sim(Ra^{*}_q)^{0.8}$, is based on the balance between the Reynolds stresses and the Ekman friction of the zonal flow at the outer boundary.

%\clearpage

\subsubsection{Ekman and Reynolds dependence}\label{sec:Ek_dependence}
\begin{figure*}[h!]%% placement specifier
\vspace{1cm}
\centering
\begin{overpic}[width=0.38\linewidth,trim={0cm 0cm 4.5cm 0cm},clip]{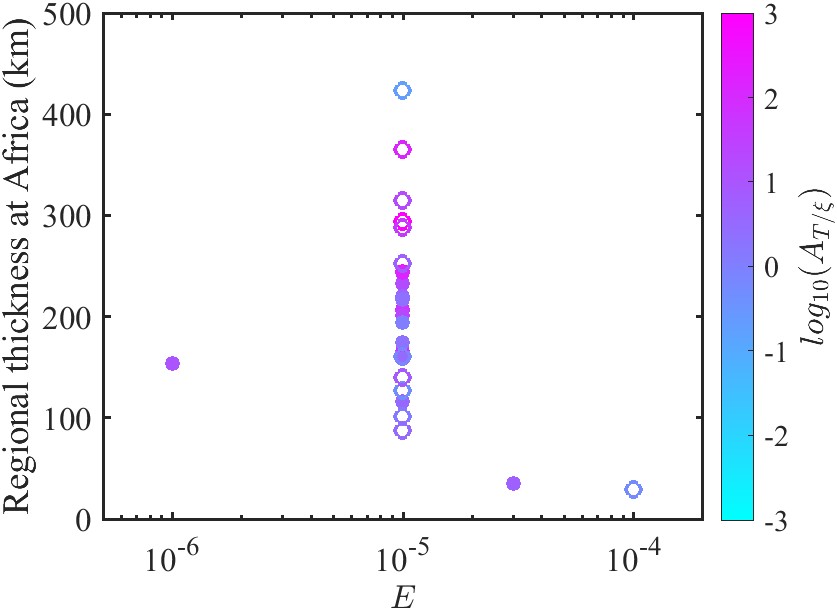}
\put(-2,86){$(a)$}
\end{overpic}
\begin{overpic}[width=0.46\linewidth,trim={0cm 0cm 0cm 0cm},clip]{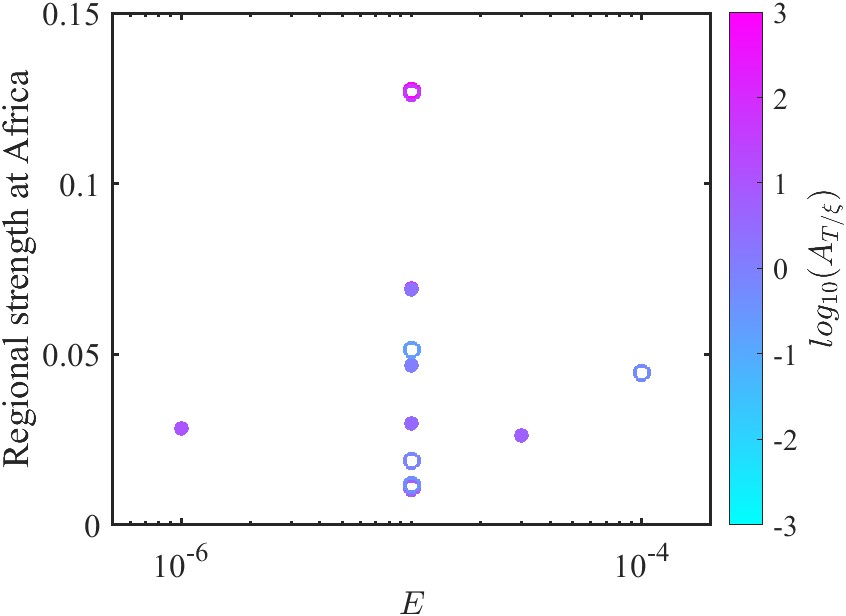}
\put(-2,72){$(b)$}
\end{overpic}
\begin{overpic}[width=0.39\linewidth,trim={0cm 0cm 4.5cm 0cm},clip]{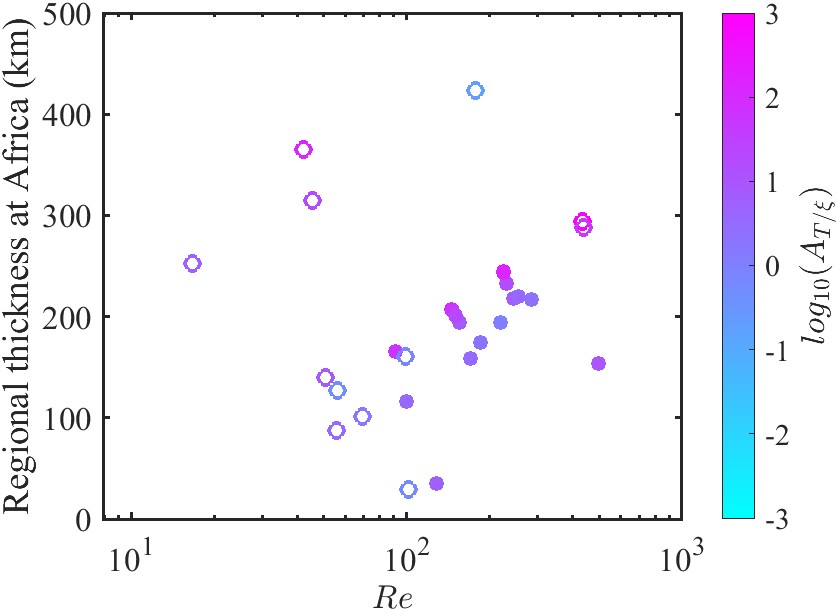}
\put(-2,86){$(c)$}
\end{overpic}
\begin{overpic}[width=0.47\linewidth,trim={0cm 0cm 0cm 0cm},clip]{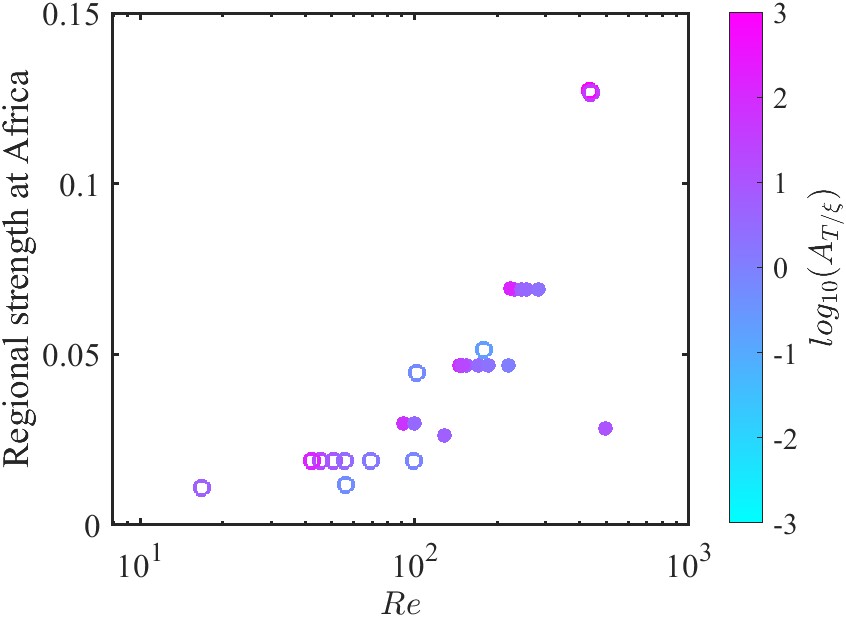}
\put(-2,72){$(d)$}
\end{overpic}
\caption{(a,b) Ekman and (c,d) Reynolds number dependence of regional (a,c) thickness and (b,d) strength of the stable region beneath Africa. The simulations with  $\FIC > 0.1$ (open symbols) are excluded from the fit in Figure~\ref{fig:scaling}.}\label{fig:Ek_dep}
\end{figure*}

 The Ekman and Reynolds number dependence of the thickness and strength of thermal RILs beneath Africa is depicted in Figure \ref{fig:Ek_dep}. The thermal RIL beneath the Pacific exhibits similar behaviour.

\section{Simulation Diagnostics}\label{app:diagnostics}
Additional data is available in the supplementary file \href{https://shorturl.at/4OxGo}{\texttt{supplementary\_data.xlsx}} included with this submission.

%\clearpage
%\bibliographystyle{elsarticle-num-names} 
%\bibliography{elsarticle}

%\end{document}

\section{Table of critical Parameters}\label{app:Rac}
\setcounter{table}{0}
\renewcommand{\thetable}{C\arabic{table}}
\begin{table}[h!]
    \centering
    \begin{tabular}{ccccc}
         $E$ & $\critRaT$ & $\critRaC$ & $\azwaveT$ & $\azwaveC$\\
         $10^{-4}$&  $16.4$&  $18.3$&  5& 1\\
         $3\times10^{-5}$&  $20.1$&  $25.6$&  8& 2\\
         $10^{-5}$&  $24.7$&  $34.1$&  12& 2\\
 $10^{-6}$& $41.0$& $70.1$& 25&28\\ 
    \end{tabular}
    \caption{Critical parameters at the onset of pure thermal convection with $\PrT=1$ and  $\boldsymbol{Q}_{T,i}=\boldsymbol{Q}_{T,o}$ , pure chemical convection with $\PrC=10$, and $\boldsymbol{Q}_{\xi,o}=0$, at various Ekman numbers. The critical Rayleigh numbers ($\critRaT$ and $\critRaC$) and azimuthal wavenumbers ($\azwaveT$ and $\azwaveC$) are calculated using SINGE \citep{kaplan_2017,vidal_2015,schaeffer_2013}.}
    \label{tab:RaC}
\end{table}

%A voir : 
%\begin{verbatim}
%A documentation about citing softwares can
%be found there: https://www.softwareheritage.org/2020/
%05/26/citing-software-with-style/ and a documentation
%about linking your article to a software here: https://
%doc.episciences.org/software/#linking-software.
%\end{verbatim}

%\clearpage

\printbibliography%[filter=onlymain]

\end{document}